\documentclass[twocolumn]{aastex631}
\usepackage{natbib}
\usepackage{graphicx}
\usepackage[flushleft]{threeparttable}
\usepackage{mathtools}
\usepackage{subcaption,caption} %
\usepackage{amsmath} 
\usepackage{enumerate}
\usepackage{enumitem}
\usepackage[normalem]{ulem} 

\shorttitle{comparison of group catalogs}
\shortauthors{Chen et al.}

\begin{document}
\title{A Comparative Study of Halo Mass Estimates from Group Catalogs and Lensing Signals}
\author[0009-0009-8953-1839]{Xinyue Chen}
\thanks{chenxinyue@bao.ac.cn}
\affiliation{National Astronomical Observatories (NAOC), Chinese Academy of Sciences, Beijing 100101, China}
\affiliation{School of Astronomy and Space Science, University of Chinese Academy of Science, Beijing 100049, China}
\author[0000-0002-9587-6683]{Weiwei Xu}
\thanks{wwxu@pku.edu.cn}
\affiliation{National Astronomical Observatories (NAOC), Chinese Academy of Sciences, Beijing 100101, China}
\affiliation{School of Physics and Astronomy, Beijing Normal University,  Beijing 100875, China}
\author[0000-0003-3899-0612]{Ran Li}
\affiliation{School of Physics and Astronomy, Beijing Normal University,  Beijing 100875, China}
\affiliation{School of Astronomy and Space Science, University of Chinese Academy of Science, Beijing 100049, China}
\author[0000-0001-8534-837X]{Huanyuan Shan}
\affiliation{Shanghai Astronomical Observatory (SHAO), Nandan Road 80, Shanghai 200030, China}
\affiliation{Key Laboratory of Radio Astronomy and Technology, Chinese Academy of Sciences, A20 Datun Road, Chaoyang District, Beijing, 100101, P. R. China}
\affiliation{University of Chinese Academy of Sciences, Beijing 100049, China}
\author[0000-0002-7336-2796]{Ji Yao}
\affiliation{Shanghai Astronomical Observatory (SHAO), Nandan Road 80, Shanghai 200030, China}
\author{Chunxiang Wang}
\affiliation{School of Electronic Science and Engineering, Chongqing University of Posts and Telecommunications, Chongqing 400065, P.R.China}

\begin{abstract}
We compare halo mass estimates from three galaxy group catalogs (redMaPPer, Yang21, and Zou21) with those derived from gravitational lensing measurements. Each catalog employs distinct methodologies, including mass-richness relations, abundance matching, and luminosity-based calibration. A linear correlation is observed between catalog-estimated and lensing-derived masses. The redMaPPer catalog shows the best agreement, especially for lower-redshift groups, with minor deviations in higher-redshift bins. Yang21 is the only catalog containing low mass groups, which gives a reasonably good mass estimation, except for the lowest mass bin. Cross-matched groups between redMaPPer and Yang21 reveal the former catalog provides more accurate mass estimation, while the Yang21 makes under-estimation of halo mass for those sharing the central galaxy with redMaPPer and over-estimation of halo mass for those with different center determination with redMaPPer and for the unique Yang21 groups. These findings emphasize the importance of redshift-dependent calibration and refined group definitions for accurate mass estimation.

\end{abstract}

\section{Introduction}

In the framework of modern cold dark matter (CDM) cosmology, the growth of dark matter structures follows a hierarchical clustering pattern \citep{white1978core,frenk2012dark}. In the early universe, small dark matter halos formed first and gradually grew through mergers and accretion. Today, the largest dark matter halos in the universe host galaxy clusters, which represent the most massive virialized structures, growing at the intersections of the cosmic web. The abundance of galaxy clusters, or the halo mass function, is crucial for constraining cosmic evolution and cosmological models (e.g., \citealt{Allen2011, abbott2020dark, abdullah2020cosmological, Chiu2023, Ghirardini2024}). Precise measurements of galaxy cluster masses, their matter distribution, and statistical properties provide a key foundation for the research of galaxy evolution \citep{mcclintock2019dark,pratt2019galaxy,umetsu2020cluster} and can put stringent constraints on the cosmological models \citep{vikhlinin2009chandra, de2016cosmological, Lesci2022}.

However, the precise measurement of massive dark matter halos first relies on accurately identifying them. Due to the invisibility of dark matter, galaxy groups or clusters are commonly used as tracers. As early as the Abell galaxy cluster era, \citet{Abell1958, Abell1989, zwicky1968catalogue} visually identified galaxy concentrations in survey data, producing one of the earliest catalogs of galaxy clusters.

In recent decades, large-scale surveys, such as the 2-degree Field Galaxy Redshift Survey \citep{colless20012df}, the Sloan Digital Sky Survey (SDSS; \citealt{york2000sloan,ahumada202016th}), and the Dark Energy Spectroscopic Instrument (DESI) Legacy Imaging Surveys \citep{blum2016decam,dey2019overview}, have provided extensive data for detecting galaxy groups and clusters.

Using data from these modern surveys, galaxy groups, and clusters have been identified through various methods. For instance, \citet{yang2005galaxy} employed the friends-of-friends (FoF) algorithm to construct a comprehensive catalog of galaxy groups, spanning a wide mass range from small groups to massive clusters. Based on photometric galaxy surveys, redMaPPer algorithm \citep{rykoff2014redmapper} produced catalogs of massive galaxy clusters \citep{rykoff2016redmapper}, identifying clusters from the SDSS DR8 database by detecting the spatial overdensity of red galaxies. Additionally, \cite{zou2021galaxy} and \cite{wen2012catalog, Wen2015, Wen2018} identified clusters by detecting galaxy number overdensity, using X-ray or Sunyaev-Zel'dovich (SZ) observations to calibrate the mass of the groups.

Once optical cluster catalogs are generated, follow-up gravitational lensing and X-ray observations play a critical role in confirming clusters, determining their masses, studying their internal structures, and analyzing their clustering properties \citep[e.g.,][]{li2009modelling,2013MNRAS.430.3359L,cacciato2009galaxy,arnaud2010universal,Umetsu2014,Bykov2015,Bellagamba2019,Tomooka2020,wang2018satellite, wang2024assessing, Xu2021,2024A&A...691A.300X}. Weak gravitational lensing, X-ray brightness, the SZ effect, and abundance matching are commonly used to calibrate the masses of galaxy clusters, providing essential inputs for studies in galaxy evolution and cosmology.

Despite these advancements, intrinsic differences remain among galaxy clusters identified using different methods. For example, the redMaPPer algorithm \citep{rykoff2014redmapper} identifies clusters based solely on red-sequence galaxies, while \citet{yang2005galaxy} and \citet{zou2021galaxy} incorporate both red and blue galaxies. These methodological differences raise questions about consistency in cluster identification across the same sky regions and the accuracy of their mass estimates. Addressing these uncertainties is crucial for ensuring reliable applications in cosmological and astrophysical studies.

In this paper, we use weak gravitational lensing data from Data Release 8 (DR8) of the DECam Legacy Survey (DECaLS) to compare the accuracy of mass estimates for three catalogs. It is important to emphasize that our objective is not to identify the best catalog but rather to explore the characteristics of the different methods used for cluster identification.

Throughout the paper, we assume the parameter values for the $\Lambda$CDM cosmology from \citep{collaboration2020planck}: the Hubble constant, H$_0$ = 67.4 km s$^{-1}$ Mpc$^{-1}$; the baryon density parameter, $\mbox{$\Omega_{\rm b}$}h^2$ = 0.0224; the cold dark matter density parameter, $\mbox{$\Omega_{\rm cdm}$}h^2$ = 0.120; the matter fluctuation amplitude, $\mbox{$\sigma_{\rm 8}$}$ = 0.811; the power index of the primordial power spectrum, n$_{\rm s}$ = 0.965; the matter density parameter, $\mbox{$\Omega_{\rm m}$}$ = 0.315. 
Since the main difference between clusters and groups is the number of galaxies within them, this work uses “group” in the following text for both of them, except for specific uses.

The organization of this paper is as follows. In Sec.~\ref{sec:data}, we describe the data, including the source catalog and the lens catalogs. Sec.~\ref{sec:method} presents the method to measure and fit the signals. Sec.~\ref{sec:result} provides the results. In Sec.~\ref{sec:conclusion}, we summarize the findings and the conclusion. 

\section{Data}
\label{sec:data}

\subsection{Source Catalog}
\label{sec:source}

\begin{figure}[t]
\centering
\includegraphics[width=0.48\textwidth]{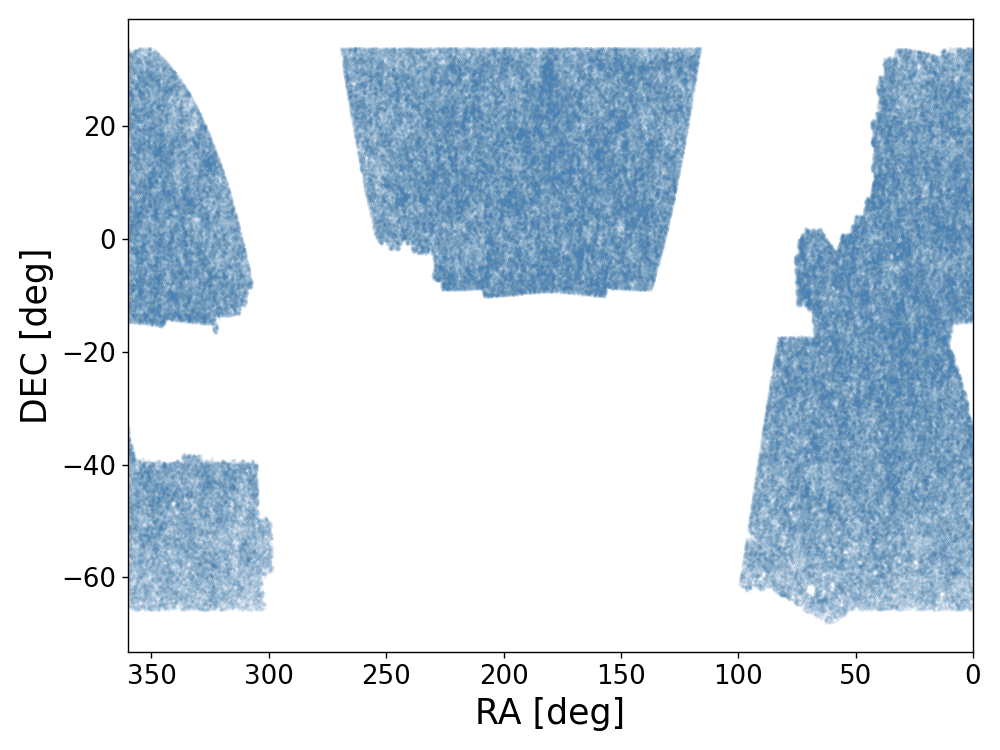}
\caption{The sky coverage of galaxies from DECaLS DR8.}
\label{fig:source} 
\end{figure}

The source galaxies used in our measurement are extracted from DECaLS DR8, which is part of Dark Energy Spectroscopic Instrument (DESI) Legacy Imaging Survey \citep{blum2016decam,dey2019overview}. DECaLS DR8 covers approximately 9500 deg$^2$ in \textit{grz} bands, as is shown in Fig.~\ref{fig:source}. In the DECaLS DR8 catalog, the sources from the Tractor catalog \citep{lang2016wise} are categorized into five different morphologies: point sources (PSF), round exponential galaxies with a variable radius (SIMP), DeVaucouleurs (DEV), exponential (EXP), and the composite (COMP) model. We retain sources above the 6\mbox{$\sigma$} detection limit in any stack as candidates. The galaxy ellipticities are estimated with parameters of the SIMP, DEV, EXP, and COMP models through the joint fitting of the three optical \textit{grz} bands. And, the multiplicative and additive biases of the DECaLS sources are estimated by the cross-matching with external shear measurements \citep{phriksee2020weak,yao2020unveiling,zu2021does}, such as Canada–France–Hawaii Telescope Stripe 82 \citep{moraes2014cfht}, the Dark Energy Survey \citep{dark2016dark}, and the Kilo-Degree Survey \citep{hildebrandt2017kids}. The residual multiplicative bias in the DECaLS DR8 shear catalog remains at $m \sim 5\%$ \citep{yao2020unveiling,phriksee2020weak}, likely due to selection effects where matched galaxies do not fully represent the observational sample (e.g., magnitude, color, size) or by small differences between simulated data and observations \citep{yao2020unveiling,2021ApJ...908...93L}. We note that our shear estimation method is comparatively rudimentary than those of large collaborations like DES required an enormous image simulation campaign to determine the multiplicative bias to a high precision (e.g., \citealt{gatti2021dark,anbajagane2025decade}). Accordingly, we adopt a conservative prior on the multiplicative shear bias $m$, as described in Sec.~\ref{sec:Am}.

We use the photometric redshifts obtained through the $k$-nearest-neighbor (kNN) method described in \citet{zou2019photometric}. This approach requires determining the linear regression relationship between the spectroscopic redshift and Spectral Energy Distribution (SED) of these $k$-nearest-neighbor galaxies. Photometric redshift is obtained from five photometric bands: 
three optical bands ($g$, $r$, and $z$) and two infrared bands ($W1$ and $W2$) from the Wide-Field Infrared Survey Explorer (WISE) survey. We have exclusively chosen samples with \mbox{$r<23$} mag and \mbox{$z<21$} mag, leading to a photometric sample of 
approximately 66 million galaxies. The final catalog exhibits a redshift bias of \mbox{$\Delta$}$z_{\rm norm}$ = 
$2.4 \times 10^{-4}$ with an accuracy of $\sigma_{\Delta z,\rm norm} = 0.017$ and an outlier rate of approximately $5.1\%$ \citep{zou2021galaxy}.

\subsection{Lens Catalogs}
\label{sec:lens}

Three catalogs of galaxy groups are compared in this work: RM sample from \citet{rykoff2016redmapper}, Y21 from \citet{yang2021extended}, and Z21 from \citet{zou2021galaxy}. Groups in these catalogs are identified with different group finders and they demonstrate distinctive characteristics. These groups are used as lenses in our galaxy-galaxy lensing ($gglens$) analysis. In the following subsections, we introduce these three catalogs in sequence.

\subsubsection{RM: Red Sequenced-based group sample}

We use the group catalog\footnote{\url{https://vizier.cds.unistra.fr/viz-bin/VizieR?-source=J/ApJS/224/1&-to=2}} from \citet{rykoff2016redmapper} (hereafter referred to as RM) identified by the red-sequence-based photometric cluster finding algorithm (redMaPPer) based on the SDSS DR8 data. 
The redMaPPer algorithm is described in detail in \citet{rykoff2014redmapper}. In short, the group finder identifies galaxy groups by pinpointing over-densities of red-sequence galaxies using three key parameters: position, luminosity, and color. 
And the richness ($\lambda$) is calculated by summing up the probability of membership over all galaxies within a certain radius. Without a halo mass estimate provided, richness is also used as a proxy for halo mass \citep{rykoff2012robust}. 

The RM catalog lists 26,111 galaxy groups
with redshifts ranging from 0.08 to 0.60. The algorithm also calculates photometric redshifts of groups, achieving unbiased performance with a precision of $\sigma_{z}/(1+z) \leq 0.01$. The accuracy of center determination within the SDSS galaxy cluster catalog is about 86\% \citep{rykoff2016redmapper}.

\begin{figure*}
\centering
    \begin{subfigure}{0.5\textwidth}
    \includegraphics[width=1\linewidth]{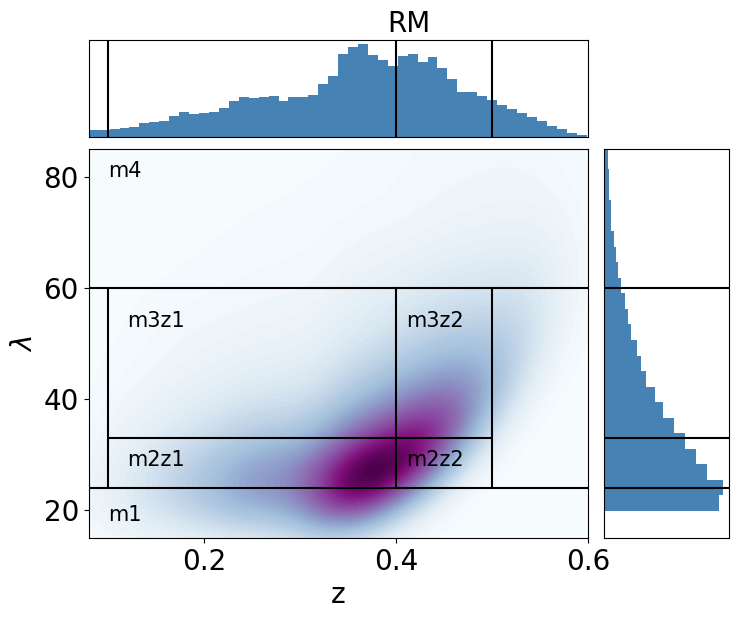}
    \label{fig:Y21-RM}
  \end{subfigure}
  \begin{subfigure}{0.49\textwidth}
    \includegraphics[width=1\linewidth]{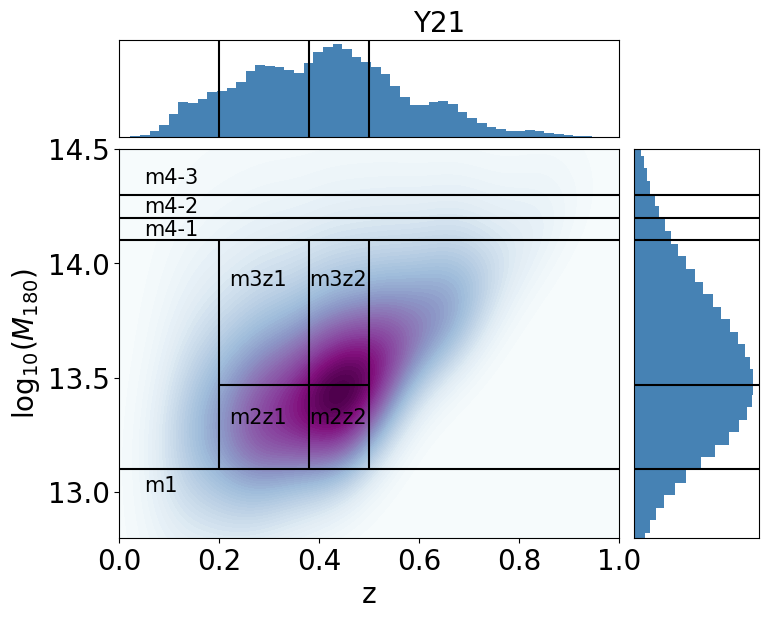}
    \label{fig:yang-red(m)}
  \end{subfigure}
  \begin{subfigure}{0.49\textwidth}
    \includegraphics[width=1\linewidth]{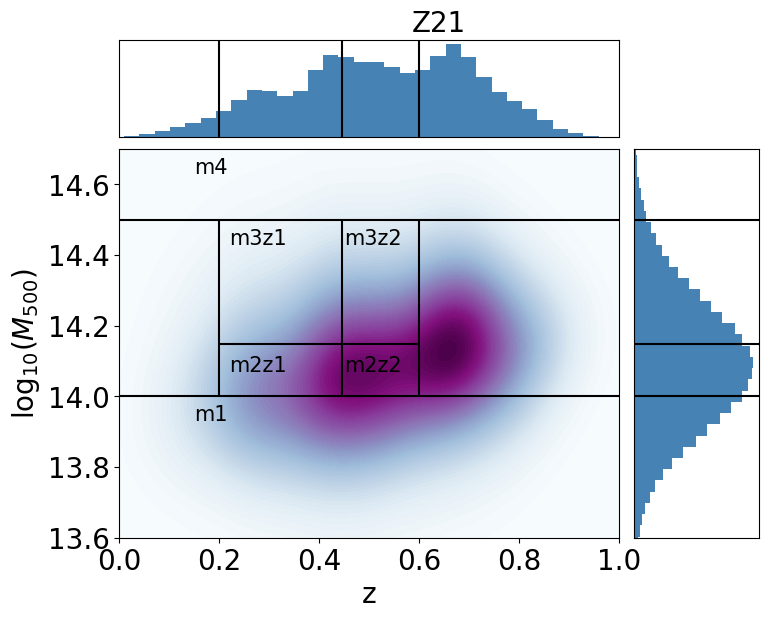}
    \label{fig:yang-red(m1)}
  \end{subfigure}
 \caption{The distribution of RM, Y21, and Z21 samples in the redshift and mass proxy parameter space. In the main panels, the number density color maps are used to highlight the densest regions. 
The shown redshift and mass proxy range are truncated to focus on the high-density region.
 Black vertical and horizontal lines are the thresholds of bins, as listed in the first part of Table~\ref{tab:bin}, and the bin names are also overlaid to specific regions. 
In the upper and right sub-panels, the histograms of redshift and mass proxy are shown, respectively. 
The mass proxy is the richness for RM, log$_{10}(M_{\rm 180}[M_{\rm \odot}h^{-1}])$ for Y21 groups, and log$_{10}(M_{\rm 500}[M_{\rm \odot}])$ for Z21 groups. }
\label{fig:distribution} 
\end{figure*}

To explore the properties of dark matter halo, we divide the RM sample into four mass bins (m1, m2, m3, m4) according to their richness. For the middle two mass bins, we further divide each of them into two redshift bins (m2$z$1, m2$z$2, m3$z$1, m3$z$2). In total, we have six subsamples with comparable size, as shown in Fig.~\ref{fig:distribution} and Table~\ref{tab:bin}.

\begin{table*}
\centering
\caption{The bin criteria of catalogs. }
\begin{threeparttable}
\begin{tabular}{cc|cccc}
\hline\hline
Catalog/Sample & Bin & $z$ & Mass Proxy$^*$& Number \\
\hline
RM$^+$ & RM(m1)  & 0.08-0.60 &$<24$     &4,575 \\
& RM(m2$z$1) &0.10-0.40 &24-33  &5,945 \\
& RM(m2$z$2) &0.40-0.50 &24-33  &2,250 \\
& RM(m3$z$1) &0.10-0.40 &33-60&4,395\\
& RM(m3$z$2) &0.40-0.50 &33-60&4,772 \\
& RM(m4)    & 0.08-0.60 &$>60.00$     &2,856 \\
\hline
Y21 & Y21(m1) &0.00-1.00 &$<13.10$    &131,945\\
& Y21(m2$z$1) &0.20-0.38 &13.10-13.47&179,872 \\
& Y21(m2$z$2) &0.38-0.50 &13.10-13.47&157,080\\
& Y21(m3$z$1) &0.20-0.38 &13.47-14.10&157,278\\
& Y2(m3$z$2) &0.38-0.50 &13.47-14.10&180,630 \\
&Y21(m4-1) &0.00$-$1.00 &14.10$-$14.20 &39,056 \\
 & Y21(m4-2) &0.00$-$1.00 &14.20$-$14.40 &45,113\\
  & Y21(m4-3) &0.00$-$1.00&$>$14.40 &24,738\\
\hline
 Z21& Z21(m1) &0.00-1.00 &$<$14.00    &170,629 \\
& Z21(m2$z$1) &0.20-0.45 &14.00-14.15&48,352\\
& Z21(m2$z$2) &0.45-0.60 &14.00-14.15&45,698\\
& Z21(m3$z$1) &0.20-0.45 &14.15-14.50&45,721 \\
& Z21(m3$z$2) &0.45-0.60 &14.15-14.50&48,377\\
& Z21(m4)     &0.00-1.00 &$>$14.50    & 12,588 \\
\hline
\hline
Y21-RM(cen.)& RM(cen.) & 0.08-0.60  &13.90-15.10& 7,214  \\
& Y21(cen.) & 0.08-0.60  &12.90-15.40    &7,214\\
\hline
Y21-RM(mem.) & RM(mem.) &0.08-0.60 &13.90-15.30&25,123 \\
& Y21(mem.) &0.08-0.60 &12.30-15.40& 25,123  \\
\hline
Y21-RM(mem. sub1) & RM(mem. sub1) &0.08-0.60 &13.90-15.10&6,393  \\
& Y21(mem. sub1) &0.08-0.60 &12.90-15.30& 6,393   \\
\hline
Y21-RM(mem. sub2) & RM(mem. sub2) &0.08-0.60 &13.90-15.30& 880 \\
& Y21(mem. sub2) &0.08-0.61 &13.60-15.30& 880  \\
\hline
unmatched samples & RM(un.) &0.08-0.60 &$>$13.90& 7,920  \\
& Y21(un.) &0.00-1.00 &$>$14.00& 33,383 \\
\hline
\hline
\end{tabular}
\begin{tablenotes}[flushleft] 
\item \textbf{Note}:
In the first part, we list the subsamples for RM, Y21, and Z21, respectively. The second part lists the subsamples based on the cross-matching result between RM and Y21. In columns, the catalog (sample) name, bin name, redshift threshold, mass proxy threshold, and halo number are shown in sequence for each sub-sample. 
The distribution in the redshift and mass proxy parameter space of subsamples in the first part of this table are shown in Fig.~\ref{fig:distribution}. The details of subsamples in the second part of this table are described in Sec.~\ref{sec:cross-matcing_samples}, Sec.~\ref{sec:crosscompare_RM_Y21}, and Sec.~\ref{sec:Mcat_Mlens}.
\item [+] Note that some groups in the RM catalog are located outside the sky coverage of the shear catalog, and do not contribute to the observed signal \citep{2010ApJ...709...97L}.
\item [*] The parameters used for mass proxy for the first part are the same as the figure note of Fig.~\ref{fig:distribution}, while the mass proxy for the second part is always log$_{10}(M_{\rm 200c}[M_{\rm \odot}h^{-1}])$.
\end{tablenotes}
\end{threeparttable}
\label{tab:bin} 
\end{table*}

\subsubsection{Y21: groups from \citet{yang2021extended}}

The group sample\footnote{\url{https://gax.sjtu.edu.cn/data/DESI.html}} identified in the work of \citet{yang2021extended} (hereafter as Y21) is also used in our measurement. These groups are identified with photometric data of the DESI Legacy Imaging Survey DR9, using a self-calibrated friends-of-friends algorithm \citep{yang2005halo,yang2007galaxy} in position and redshift space.

The method, as described by \citet{yang2005halo,yang2007galaxy}, first assigns galaxies to potential groups based on two specially defined linking lengths: one in the line-of-sight direction ($l_{z}$) and the other in the transverse direction ($l_{\rm p}$). After this initial grouping, the luminosity-weighted centers of candidate groups are identified. The total luminosity of each group is then measured to estimate the group’s mass using the mass-to-light relation.

Galaxies are considered members of a group if they meet a specific three-dimensional density contrast criterion. This involves selecting galaxies within a radius of $R_{\rm 180}$ around the group center. $R_{\rm 180}$ is defined as the radius at which the dark matter halo exhibits an overdensity of 180 times greater than the universe’s background density. The algorithm iterates this process 3-4 times until both the mass-to-light ratio and the group memberships converge.
This algorithm has been successfully applied to both photometric and spectroscopic redshift data, resulting in high purity and completeness for the detected galaxy groups \citep{yang2005halo,yang2007galaxy,yang2021extended}.

The Y21 catalog, derived from photometric data from the DESI Legacy Imaging Survey DR9, covers both the Southern and Northern Galactic Caps, with a redshift range of $0.0 \leq z \leq 1.0$. Member galaxies in the Y21 catalog have photometric redshifts estimated using a random forest algorithm from the Photometric Redshifts for the Legacy Surveys (PRLS) catalog \citep{2021MNRAS.501.3309Z}, with additional quality control criteria applied \citep{yang2021extended}. Whenever available, spectroscopic redshifts are used to replace the photometric estimates, incorporating data from various catalogs including BOSS, SDSS, WiggleZ, GAMA, COSMOS2015, VIPERS, eBOSS, DEEP2, AGES, 2dFLenS, VVDS, OzDES, 2MASS Redshift Survey, 6dF Galaxy Survey Data Release 3, and the 2dF Galaxy Redshift Survey.

We follow \citet{2022MNRAS.511.3548S} in this work, selecting groups with a richness $\lambda > 5$. After applying this richness criterion, the final sample consists of 1,327,005 groups. We take the position of the most massive member as the halo center for each group. Then, the Y21 groups with $\lambda > 5$ are divided into 6 bins (m1, m2$z$1, m2$z$2, m3$z$1, m3$z$2, m4) similarly with RM. Focusing more on massive groups, we further divide the most massive bin (m4) into three high-mass subsamples(m4-1, m4-2, and m4-3), as summarized in Table~\ref{tab:bin} and shown in Fig.~\ref{fig:distribution}.

\subsubsection{Z21: Groups identified by \citet{zou2021galaxy}}

We also utilize another group catalog from \citet{zou2021galaxy} (hereafter referred to as Z21). In this catalog, groups are identified from DESI DR8 data using the Fast Search and Find of Density Peaks (CFSFDP) algorithm \citep{rodriguez2014clustering}. The groups are detected by locating overdensities of galaxies within specific redshift slices. The Z21 catalog includes 540,432 groups spanning redshifts from 0.0 to 1.0, with a reported false detection rate of approximately 3.1\% \citep{zou2021galaxy}.

The halo masses in the Z21 catalog are estimated using the total $r$-band luminosity of member galaxies within a 1 Mpc radius, referred to as \(L_{\rm 1Mpc}\). The relationship between \(L_{\rm 1Mpc}\) and halo mass was established based on a calibration sample of galaxy groups with corresponding X-ray or SZ observations (hereafter as `Zou(X-ray/SZ)' sample), where the masses were independently measured. This relationship is then applied to the whole Z21 catalog for mass estimation. In this work, the Z21 catalog is divided into six bins based on specific redshift and mass thresholds, following similar criteria to that of the RM catalog, as illustrated in Fig.~\ref{fig:distribution}, and detailed in Table~\ref{tab:bin}.

\subsection{Mass distribution of halos in different catalogs}

\begin{figure}
\centering
\includegraphics[width=0.45\textwidth]{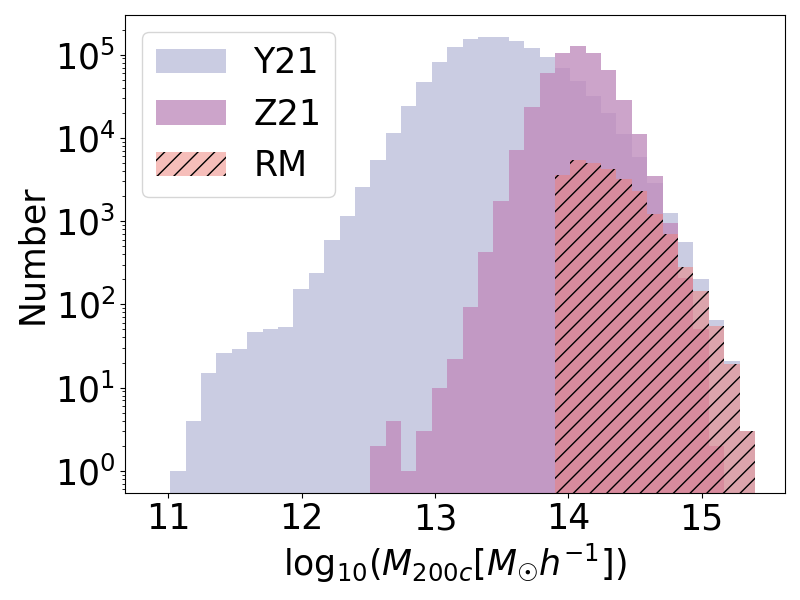}
\caption{Histogram of halo mass in RM, Y21, and Z21 catalogs.}
\label{fig:masshis}
\end{figure}

Fig.~\ref{fig:masshis} shows the halo mass distribution across different galaxy group catalogs, revealing notable differences. The detailed mass conversion method is described in Sec.~\ref{sec:Mcat_Mlens}. The RM catalog predominantly includes high-mass galaxy groups. In lower-mass groups, the correlation between galaxy brightness and color becomes weak, reducing the effectiveness of the redMaPPer method, which relies on the red-sequence relationship. In the RM catalog, the richness is $\lambda > 19$, resulting in a sharp cutoff in the distribution at the low-mass end. In contrast, the Y21 catalog, which is based on the friends-of-friends and overdensity methods, includes both high-mass groups and low-mass systems, with a peak at $\sim 10^{13.4} M_{\odot}/h$. The Z21 catalog falls between the two, encompassing a range of masses with intermediate characteristics.

\subsection{Cross-matched group samples}
\label{sec:cross-matcing_samples}

To compare mass estimates across different galaxy group catalogs, we cross-match galaxy groups between RM and Y21 catalogs. A natural approach to matching requires the groups to be located at the same position and redshift, with the same central galaxy. We require the central galaxy position offset between the two catalogs to be less than 3$\arcsec$, and the redshift difference to be less than 0.01. Ultimately, this produces a cross-matched group sample ‘Y21-RM(cen.)’, containing 7,214 groups. For this set of groups, their characteristics parameters from the RM catalog are named as ‘RM(cen.)' samples, and parameters from the Y21 catalog are named as ‘Y21(cen.)' samples (Table~\ref{tab:bin}).

However, in many cases, even though the same galaxy groups are found in both catalogs, the definition of the central galaxy differs. To address this, we apply an alternative two-sided cross-matching method. In this approach, for a galaxy group A in Catalog-1, we identify its member galaxies and match them to their corresponding galaxy groups in Catalog-2 (requiring a position offset of less than $3\arcsec$). The group in Catalog-2 that contains the largest number of A’s member galaxies is considered the matched group for 1-A. We repeat this process for each galaxy group in Catalog-1 and each group in Catalog-2 iteratively until the result converges, establishing a cross-matched catalog between the two datasets. 

Since the Z21 catalog does not provide member galaxy information, we apply this matching method only to the RM and Y21 catalogs. For each galaxy group in Y21, we identify its corresponding group in RM using the member galaxy cross-matching strategy. Finally, through this member galaxy cross-matching, we obtain a sample of 25,123 groups (hereafter referred to as ‘Y21-RM(mem)’).

\section{Method}
\label{sec:method}
\subsection{The lensing signal}

In this section, we briefly introduce the measurement of stacked lensing signals. The gravitational well of the halo causes image distortions of source galaxies, resulting in tangential shears ($\gamma$) with elongated shapes. 
Under the weak gravitational lensing limitation, the tangential shear of source galaxies can be characterized as “reduced shear”,
 \begin{align}
\mathbf{g}\equiv\frac{\gamma}{1-\kappa},
\end{align}
where $\kappa$ is the dimensionless surface mass density, defined as \mbox{$\kappa=\Sigma\left(R\right)/\Sigma_{\rm \mathrm{crit}}$}. 

The gravitation of groups, which is a localized mass distribution, causes a positive shear along the tangential direction relative to the center of the over-density. This net tangential shear leads to the stretching and alignment of background source galaxy images along the tangential direction. Thus, the magnitude of the azimuthally averaged tangential shear at the projected radius $R$ can be predicted based on the projected excess surface mass density along the line of sight, as
 \begin{align}
\gamma_\mathrm{T}=\frac{\overline{\Sigma}(<R)-\overline{\Sigma}\left(R\right)}{\Sigma_\mathrm{crit}}\equiv\frac{\Delta\Sigma\left(R\right)}{\Sigma_\mathrm{crit}}.
\end{align}

The $\overline{\Sigma}\left(R\right)$ is the average surface mass density at the radius of $R$, while the $\overline{\Sigma}(<R)$ is the average surface mass density within the radius of $R$. The critical surface mass density $\Sigma_\mathrm{crit}$ 
is defined as
 \begin{align}
\Sigma_{\rm \mathrm{crit}}=\frac{c^{2}}{4\pi G}\frac{D_{\rm s}}{D_{\rm l}D_{\rm ls}},
\end{align}
where $D_{\rm s}$ and $D_{\rm l} $ are the angular diameter distances to the background source galaxies and lens galaxies, respectively, and $D_{\rm ls}$ is the angular diameter distance between the lens and the source, $c$ is the velocity of light in a vacuum (constant). 

To obtain $\Delta\Sigma$, we stack lens-source pairs in comoving radial bins. To avoid the influence of redshift measurement errors and reduce any ``dilution" from foreground galaxies, which leads to mis-classification, we only consider sources that satisfy $z_{\rm s}$ $>$ $z_{\rm l} + 0.1$. Therefore, $\Delta\Sigma$ is corrected as 
\begin{equation}
\label{eq:deltasigma}
\Delta\Sigma\left(R\right)=\frac{\Sigma_{\rm ls}w_{\rm ls}\gamma_{t}^{\rm ls}\Sigma_{\mathrm{crit}}}{\Sigma_{\rm ls}w_{\rm ls}},
\end{equation}
where $\gamma_{\rm t}^{\rm ls}$ is the tangential shear and $w_{\rm ls}=w_{\rm n}·\Sigma_{\mathrm{crit}}^{-2}$, $w_{\rm n}$ is a weight factor introduced to account for intrinsic scatter $\sigma_\epsilon$ and shape measurement $\sigma_\mathrm{e}$ errors of the ellipticity \citep{dey2019overview}, defined as $w_\mathrm{n}=1/(\sigma_\epsilon^2+\sigma_\mathrm{e}^2)$ in this work. After these corrections, the final lensing signal is represented as 
 \begin{align}
\Delta\Sigma^{\mathrm{cal}}(\begin{matrix}R\end{matrix})=\frac{\Delta\Sigma(\begin{matrix}R\end{matrix})}{1+K(\begin{matrix}z_{\rm l}\end{matrix})},
\end{align}
 \begin{align}
1+K\left(z_{\rm l}\right)=\frac{\sum_{\rm ls}w_{\rm ls}\left(1+m\right)}{\sum_{\rm ls}w_{\rm ls}},
\end{align}
where $m$ is the multiplicative bias.

In this work, we utilize the \textsc{swot} software\footnote{\url{http://jeancoupon.com/swot}} \citep{coupon2012galaxy} to detect the stacked signals. It is a fast tree code that can be used for computing two-point correlations, histograms, and gglens signals in large data sets. The \textsc{swot} software can be parallelized to enhance computational efficiency. We use comoving distance projection to measure the signal and estimate the statistical error with a Jackknife resampling of 64 sub-regions with equal area, removing one subsample at a time for each Jackknife realization. The parameter settings for \textsc{swot} can be found in Table~\ref{tab:Table2}.
We restrict the radii in the range of 0.3–5 Mpc$/h$ and divide them into 10 logarithmic bins to measure the signals. 

\begin{table*}
\centering
\caption{Parameter Settings of \textsc{swot}}
\label{tab:Table2} 
\begin{tabular}{ c|c|c }
\hline\hline
Parameters & Value & Meaning\\
\hline
corr & gglens &Type of correlation \\
 range& 0.3, 5 &Correlation range (in unit of Mpc $h^{-1}$)\\
 nbins & 10 & Numbers of bins\\
   err& Jackknife &Resampling method \\
   nsub& 64& Number of resampling subvolumes \\
   H$_0$& 67.4& Hubble parameter \\
$\mbox{$\Omega_{\rm m}$}$ & 0.315& Relative matter density\\
   $\mbox{$\Omega_{\rm L}$}$ & 0.684 &Relative energy density\\
   $\Delta$ & 0.1& Minimum redshift difference between the source and the lens\\
   proj.& como& Projection \\
\hline
\hline
\end{tabular}
\end{table*}

\subsection{The lensing model}
\label{sec:The Lensing Model}

In the model, three components are considered, including the contribution from the main DM halo, a mis-centering term, and the two-halo term, as
\begin{equation}
\label{eq:model1}
 \begin{split}
\Delta\Sigma\left(R\right)=p_{\rm \mathrm{cen}}\Delta\Sigma_{\rm \mathrm{NFW}}\left(R\right)~~~~~~~~~~~~~~~\\
+\left(1-p_{\rm \mathrm{cen}}\right)\Delta\Sigma_{\rm \mathrm{NFW}}^{\mathrm{off}}\left(R|R_{\rm \mathrm{mis}}\right)\\
+\Delta\Sigma_{\rm \mathrm{2h}}\left(R\right).~~~~~~~~~~~~~~~~~~~~~
\end{split}
\end{equation}

Firstly, we use the Navarro-Frenk-White (NFW, \citealt{navarro1997universal}) profile to describe the contribution of the central dark matter halo, whose density distribution is in the form of, 
 \begin{align}
\rho\left(r\right)=\frac{\bar{\delta}\bar{\rho}}{\left(r/r_{\rm s}\right)\left(1+r/r_{\rm s}\right)^2},\label{NFW}
\end{align}
where $r$ is the three-dimensional distance to the center galaxy, $\bar{\rho}$ is the average density of the universe, and $r_{\rm s}$ is the scale radius (indicating the distribution of dark matter) where the density slope $d$ln$\rho$/$d$ln$r=-2$. $\bar{\delta}$ is the normalization, which is dependent only on the concentration. The concentration is defined as $c = r_{\rm vir}/r_{\rm s}$, and believed to be correlated with halo mass based on observations \citep{shan2017mass, Xu2021,mcclintock2019dark} and numerical simulations \citep{duffy2008dark}.

In addition, the miscentering term,  $(1-p_{\rm cen})~\Delta\Sigma_{\mathrm{NFW}}^{\mathrm{off}}(R|R_{\mathrm{mis}})$, accounts for the stacked signal of inaccurate identification of the halo center. It is characterized by a characteristic miscentering length ($R_{\rm mis}$), and the miscentering fraction as (1$-p_{\rm cen}$), where the $p_{\rm cen}$ represents the fraction of halos without miscentering effect. 
The stacking of miscentering halos can affect the detected tangential shear profile and significantly reduce the lensing signal \citep{johnston2007cross}. 

Furthermore, it is necessary to take into account the gravitational influence from nearby DM halos, represented by the two-halo term ($\Delta\Sigma_{\rm \mathrm{2h}}$), which becomes dominant at the outskirt of groups. The contribution from the two-halo term in this work is estimated using the \textsc{halofit} model \citep{Lewis1999, Lewis2002, Howlett2012}, which accounts for the nonlinear scaling of the matter power spectrum. This estimation is performed through the \textsc{camb} package\footnote{\url{https://github.com/cmbant/CAMB}}.

In the fitting model, 
there are four parameters: dark matter halo mass ($M_{\rm \text{vir}}$), concentration (\(c\)), the miscentering length ($R_{\rm mis}$), and fraction of halos without miscentering ($p_{\rm \mathrm{cen}}$). For more detailed information about our model, refer to the \textsc{Cluster\_toolkit} package\footnote{\url{https://cluster-toolkit.readthedocs.io/en/latest/}} \citep{mcclintock2019dark}.

\subsection{Systematics}
\label{sec:Systematics}

During the model fitting of weak gravitational lensing signals, in addition to considering the contributions from various components mentioned above, it is also necessary to further include the multiplicative corrections \citep{mcclintock2019dark}, such as the boost factor ($B(\theta)$, Sec.~\ref{sec:boost}), photo-$z$ dilution correction (Sec.~\ref{sec:zph_dilution}), and the shear multiplicative bias ($m$, Sec.~\ref{sec:Am}).

\subsubsection{Boost Factor}
\label{sec:boost}
We adopt the widely used estimator \citep{mandelbaum2006density, singh2017galaxy, yao2020unveiling} for computing the galaxy-galaxy lensing correlation functions ($w_{\gamma g}$), defined as:
\begin{equation}
w_{\gamma g}(\theta) = 
B(\theta)\frac{\sum\limits_{\mathrm{ED}} w_j \gamma_{+,j}}{\sum\limits_{\mathrm{ED}} w_j (1 + m_j)} - 
\frac{\sum\limits_{\mathrm{ER}} w_j \gamma_{+,j}}{\sum\limits_{\mathrm{ER}} w_j (1 + m_j)},
\label{eq:shearestimator}
\end{equation}
where the summations over ED and ER refer to the pairs between the tangential shear and the lens number density in the data lens catalog (ED) or in the random lens catalog (ER), respectively. The numerator in each term represents the stacked tangential shear weighted by the inverse variance weight \( w_j \) of the \( j \)-th source galaxy. The denominator provides normalization, accounting for both \( w_j \) and the shear multiplicative bias correction factor \( (1 + m_j) \). To account for the excess clustering of source galaxies that are physically associated with the lens but erroneously classified as background, we include the effect of the boost factor \citep{varga2019dark,mcclintock2019dark,simet2017weak,Chiu2022,bocquet2024spt,grandis2024srg,kleinebreil2024srg}, denoted as \( B(\theta) \) in Eq.~\ref{eq:boost}. The boost factor is defined as:
\begin{equation}
B(\theta) = \frac{\sum_{\mathrm{ED}} w_j (1+m_j)}{\sum_{\mathrm{ER}} w_j (1+m_j)},
\label{eq:boost}
\end{equation}
where \( \sum_{\mathrm{ED}} \) and \( \sum_{\mathrm{ER}} \) represent the weighted sum over shear-lens and shear-lens(random) pairs, respectively, and \( w_j \) is the weight assigned to each source galaxy. This factor quantifies the excess contribution from physically associated source-lens pairs, which leads to a scale-dependent bias at small angular separations due to the clustering of galaxies \citep{bernardeau1997effects, hamana2002source, yu2015source}. In particular, when the redshift distributions of source and lens galaxies overlap, the weighted sum over shear-lens pairs, \( \sum_{\mathrm{ED}} \), can be significantly contaminated by physically associated pairs arising from galaxy clustering. More details about boost factor term for our samples can be seen in Appendix~\ref{app:boost}.

\subsubsection{ Photo-$z$ Dilution Correction}
\label{sec:zph_dilution}

A key challenge in measuring the excess surface density, $\Delta\Sigma$, arises from the uncertainty in source galaxy distances, which are typically inferred from photometric redshifts \citep{2024OJAp....7E..57L}. These uncertainties affect the calculation of the critical surface density, $\Sigma_{\rm crit}$, leading to a bias in the lensing signal. For instance, galaxies that appear to lie behind a lens due to photometric redshift errors may in reality be in front of it and thus do not contribute to the lensing signal, resulting in a downward bias in the measurement.

Assuming the lens redshifts are spectroscopic, and using the estimated critical surface density $\Sigma_{\rm crit,lp}$ based on the photometric redshift of the source (instead of the true $\Sigma_{\rm crit,ls}$), the measured $\Delta\Sigma$ becomes biased relative to the true value $\Delta\Sigma_{\rm true}$. When stacking over many source-lens pairs with the weight of $w_{\rm ls}$, this bias takes the form of a multiplicative correction:
\begin{equation}
\Delta\Sigma_{\text{true}} = f_{\text{bias}} \langle \hat{\Delta\Sigma} \rangle,
\end{equation}
where
\begin{equation}
f_{\mathrm{bias}} = \frac{\sum_{\mathrm{ls}} w_{\mathrm{ls}}}{\sum_{\mathrm{ls}} w_{\mathrm{ls}} \left( \frac{\Sigma_{\mathrm{crit},\mathrm{lp}}}{\Sigma_{\mathrm{crit},\mathrm{ls}}} \right)}.
\end{equation}
This correction factor can be evaluated as a function of lens redshift using a subset of galaxies with both photometric and spectroscopic redshifts \citep{Nakajima2012}.

To obtain the spectroscopic redshifts of the shear catalog, we employ Self-Organising Maps (SOMs, \citealt{kohonen1982self}), an unsupervised machine learning technique that projects high-dimensional data onto a low-dimensional (typically 2D) grid while preserving topological structure. In our baseline setup, we train a 101$\times$101 hexagonal-cell SOM with toroidal topology using 10 colour indices and magnitudes from five photometric bands ($g$, $r$, $z$, $W1$, $W2$). This structured partitioning of colour–magnitude space enables direct association between photometric and spectroscopic sources within the same SOM cells, facilitating robust redshift distribution reconstruction and providing a powerful diagnostic for identifying systematic biases. The training set consists of the subset of DECaLS DR8 data with available spectroscopic redshifts, which we divide into five redshift bins within following redshift ranges, $0.1–0.3$, $0.3–0.5$, $0.5–0.7$, $0.7–0.9$, and $>$0.9 \citep{wright2020photometric}, comprising a total of 714,817 galaxies. For the training set, the difference between \( z_{\text{spec}} \) and \( z_{\text{SOM}} \) shows biases \( \Delta \langle z_i \rangle < 0.0035 \).

\begin{figure}
\centering
\includegraphics[width=0.5\textwidth]{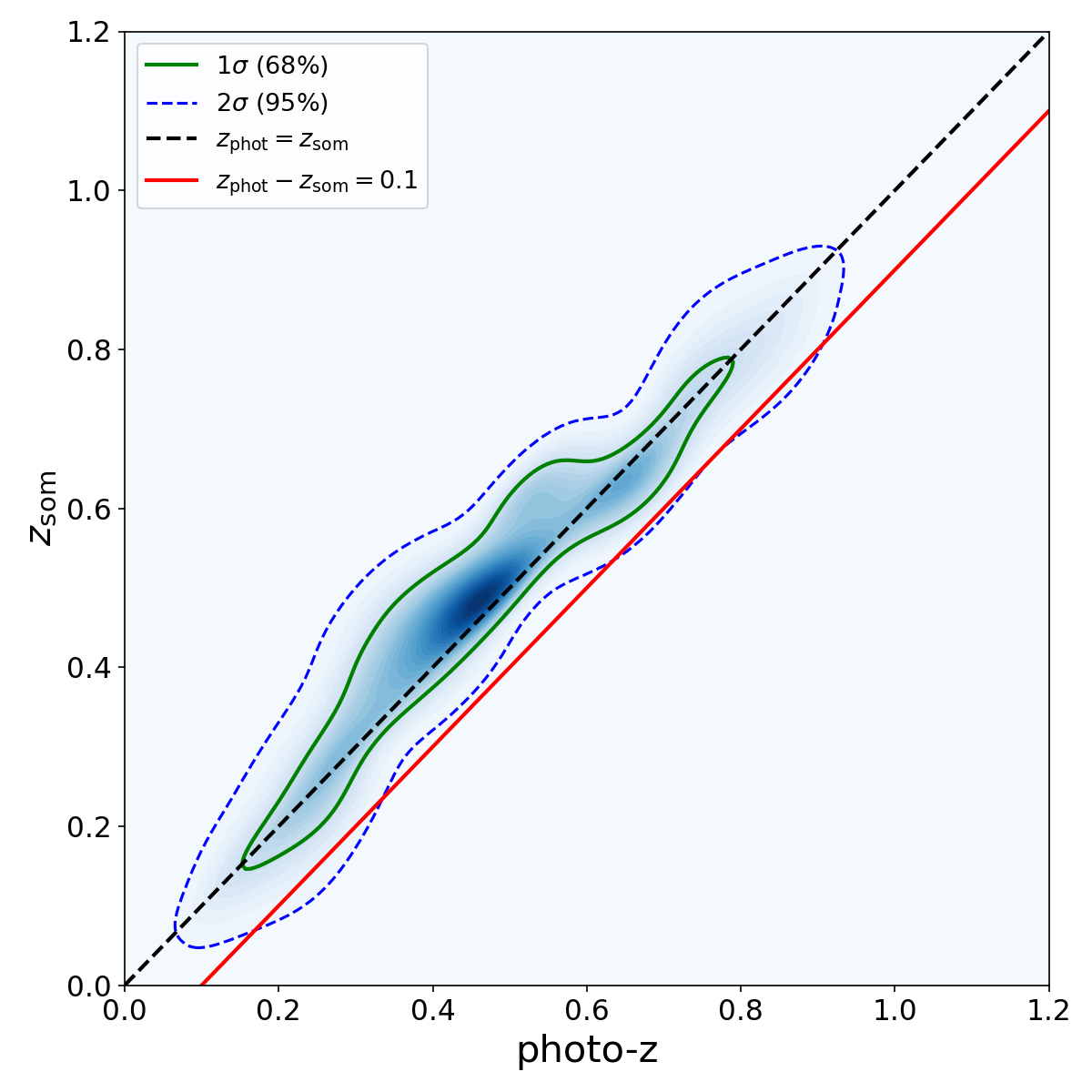}
\caption{The distribution of photo-$z$ and $z_{\text{SOM}}$ of galaxies in the shear catalog.
The red line represents the relation of $z_{\rm phot}-z_{\rm SOM} = 0.1$. The colormap and contours represent the distribution of number density.}
\label{fig:som}
\end{figure}

For those sources without spectroscopic redshifts, we obtain their \( z_{\text{SOM}} \) estimates using the method described above. Approximately 5\% of the sources cannot be assigned a prediction, the majority ($\sim 3\% $) of which originate from galaxies with photometric redshifts greater than 0.9. The comparison of photo-$z$ and $z_{\text{SOM}}$ of sources from shear catalog is shown in Fig.~\ref{fig:som}.
 \textit{photo-\(z\) dilution} effect dominates in the region with  \( z_{\mathrm{phot}} - z_{\mathrm{lens}} > 0.1 \), where galaxies (1.97\%) with true redshifts \( z_{\mathrm{SOM}} < z_{\mathrm{lens}} \) are scattered to higher photometric redshifts 
and mistakenly taken as background sources.

On the other hand, the criterion of \( z_{\mathrm{phot}} - z_{\mathrm{lens}} > 0.1 \) assess the effectiveness of the selection in removing clustering contamination (see Sec.~\ref{sec:boost}), as unlensed galaxies with \( z_{\mathrm{SOM}} \approx z_{\mathrm{lens}} \) dilute the signal and tend to lie near this relation.
The $f_{\rm bias}$ is estimated to deviate from 1 by $\sim 3\%$ in all five redshift bins.

\subsubsection{Multiplicative Shear Bias}
\label{sec:Am}

The residual multiplicative bias mentioned in Sec.~\ref{sec:source} was tested on DECaLS DR9, yielding approximately 5\% (see Fig.~\ref{fig:shear_bias}), which is consistent with \citet{phriksee2020weak}. The factor $ \mathsf{A}_{m}=1+m$  is setted as a parameter in fitting, and the $m$ is assigned a Gaussian prior of $0 \pm 5\%$. This step adjusts our model to the following form:
 \begin{align}
 \label{eq:model2}
\Delta\Sigma_{\mathrm{model}}\left(R\right)=\mathsf{A}_{m}\Delta\Sigma(R).
\end{align}

\section{Result}
\label{sec:result}

\subsection{Cross-Catalog Comparison of RM and Y21} 
\label{sec:crosscompare_RM_Y21}

\begin{figure*}
\centering
    \begin{subfigure}{0.48\textwidth}
    \includegraphics[width=1\linewidth]{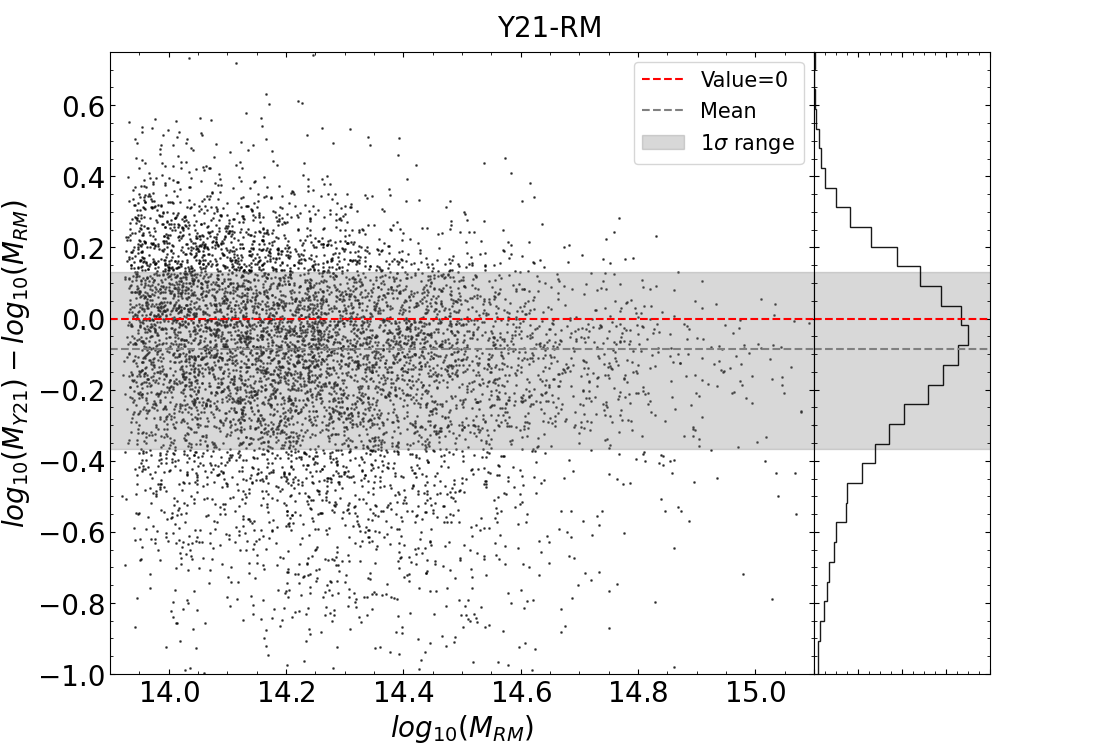}
    \label{fig:Y21-RM}
  \end{subfigure}
  \begin{subfigure}{0.48\textwidth}
    \includegraphics[width=1\linewidth]{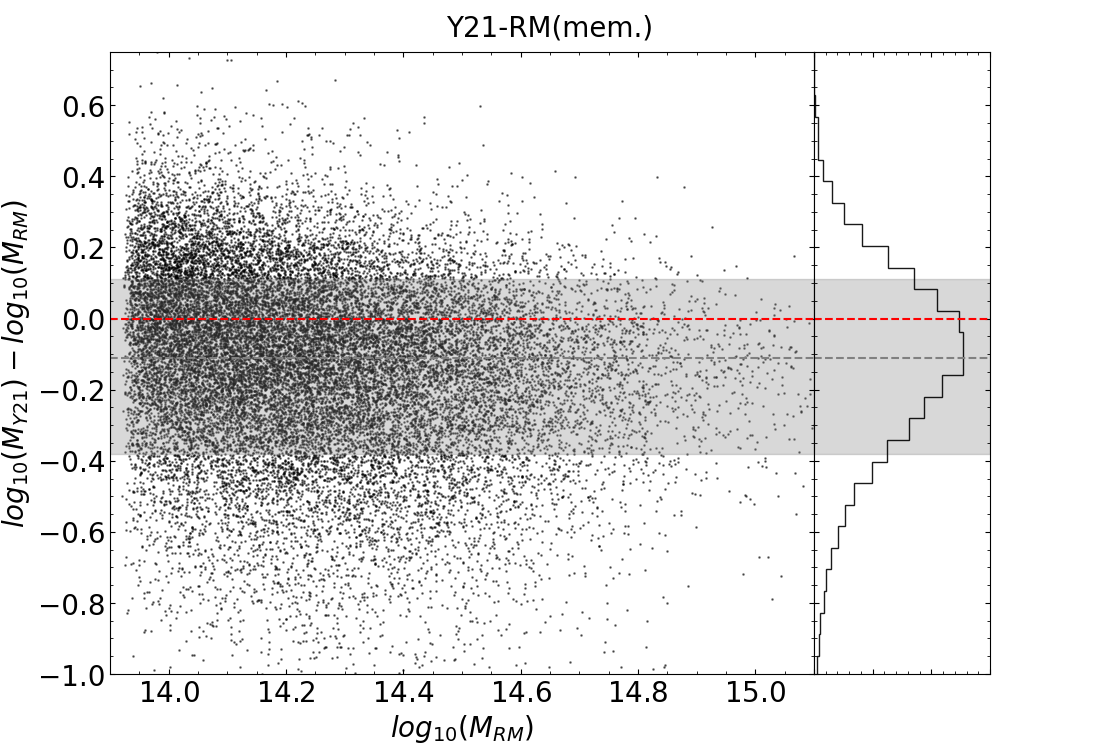}
    \label{fig:yang-red(m)}
  \end{subfigure}
  \begin{subfigure}{0.48\textwidth}
    \includegraphics[width=1\linewidth]{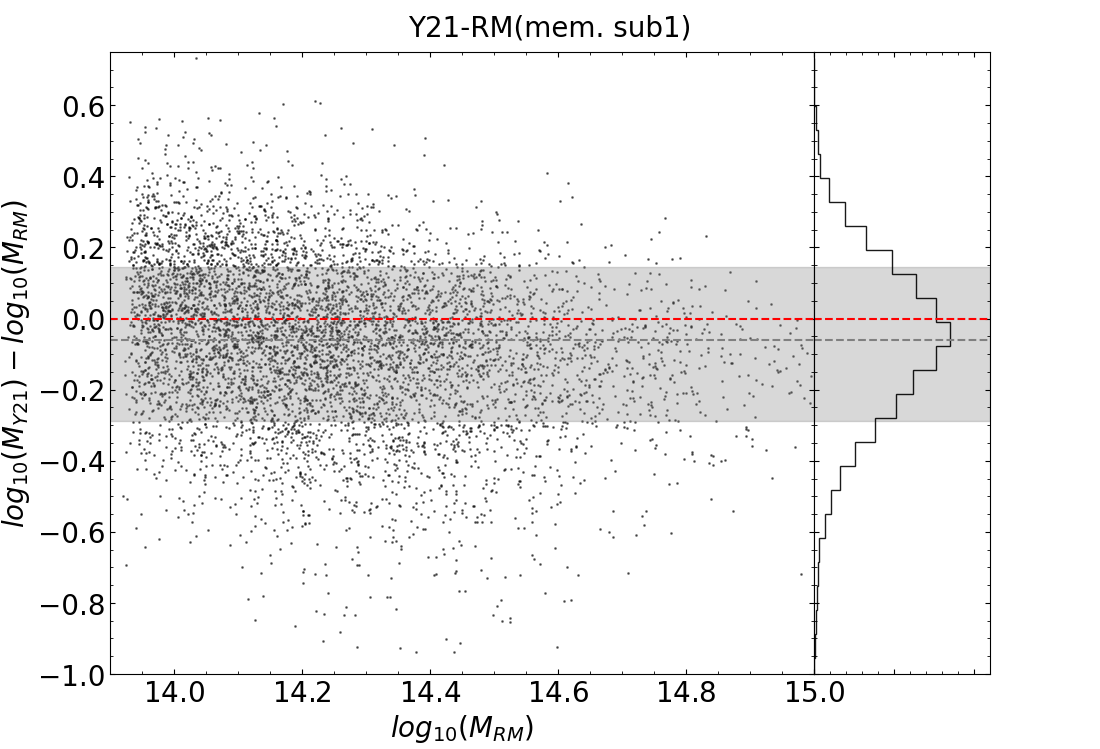}
    \label{fig:yang-red(m1)}
  \end{subfigure}
 \caption{The relation between mass difference and the halo mass in difference catalogs. In each panel, the Y-axis denotes the difference between two estimated halo masses, while the X-axis signifies the mass from RM. The horizontal grey dashed line shows the mean value of mass difference, while its $1\sigma$ range is labeled as the grey shaded area. The horizontal red dashed line shows the location with no mass difference. The distribution of mass difference is shown in the right subpanels.}
\label{fig:crosscomparison} 
\end{figure*}

For galaxies in the same sky region, different galaxy group-finding methods yield distinct group catalogs. Even for galaxy groups that include similar member galaxies, different catalogs often provide varied mass estimates. In Fig.~\ref{fig:crosscomparison}, we present a comparison of the assigned masses for groups both listed in the RM and Y21 catalogs.

Whether using the position of central galaxies as the matching criterion or employing a two-sided matching of member galaxies (as described in Sec.~\ref{sec:cross-matcing_samples}), the mean relationship between the masses assigned by Y21 and RM is in reasonable agreement, with an average difference of log$_{10}(M_{\rm vir})$ as 0.086 for `Y21-RM(cen.)' and 0.110 for `Y21-RM(mem.)', as shown in Fig.~\ref{fig:crosscomparison}. The halo mass assigned by Y21 is slightly lower than that from RM, both with a scatter of approximately 0.25 dex.

\begin{figure}
  \centering
    \includegraphics[width=0.99\linewidth]{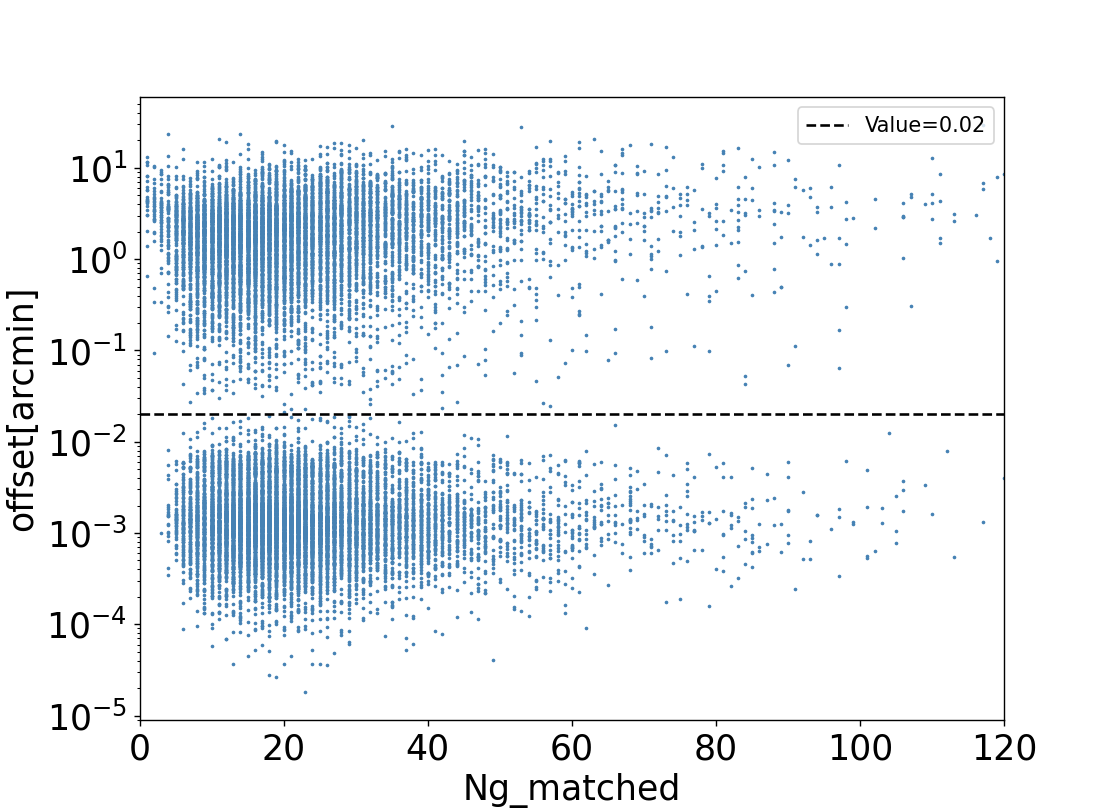}
 \caption{The distribution of center separation of `Y21-RM(mem.)' sample and the number of matched member galaxies in each group. Black dashed line is a threshold of 0.02 arcmin, which is used to separate the `Y21-RM(mem. sub1)' and `Y21-RM(mem. sub2)', as described in Sec.~\ref{sec:crosscompare_RM_Y21}.}
\label{fig:sepmember}
\end{figure}

We find that 96.2\% of RM groups can find a corresponding Y21 group through the member cross-matching. It is important to note that the member-matched sample and the center-matched sample are not entirely equivalent. In Fig.~\ref{fig:sepmember}, we show the distribution of center offsets and the number of matched members for the groups in the member cross-matched sample, `Y21-RM(mem.)'. The distribution exhibits a clear bimodality: the lower branch corresponds to groups where the central galaxy in the parent group catalogs is the same, while the upper branch represents groups where the central galaxy differs between the two catalogs.

We further select two subsamples from the cross-matched group sample: one where both catalogs assign the same center, and another where the assigned centers differ, as shown in Fig.~\ref{fig:sepmember}. For the first subsample, we require the center position offset from the parent catalog to be less than $0.02\arcmin$ and the redshift difference to be less than 0.01, resulting in 6,393 groups, denoted as `Y21-RM(mem. sub1)'. For this subsample, the mass estimates from the two catalogs are consistent, with the mean difference reduced to 0.059 and the scatter decreasing to 0.2 dex, as shown in Fig.~\ref{fig:crosscomparison}. The second subsample includes groups with different assigned centers. In this case, we further require that the two parent group catalogs identify more than 50\% of the same member galaxies in the cross-matched group. This selection results in 880 groups, referred to as `Y21-RM(mem. sub2)'.

In the next section, we will compare the estimated masses from different catalogs with lensing-measured masses to further explore the biases in different mass estimators.

\subsection{The fitting of galaxy-galaxy lensing signals}
\label{sec:fitting}

\begin{figure}
  \centering
    \includegraphics[width=0.99\linewidth]{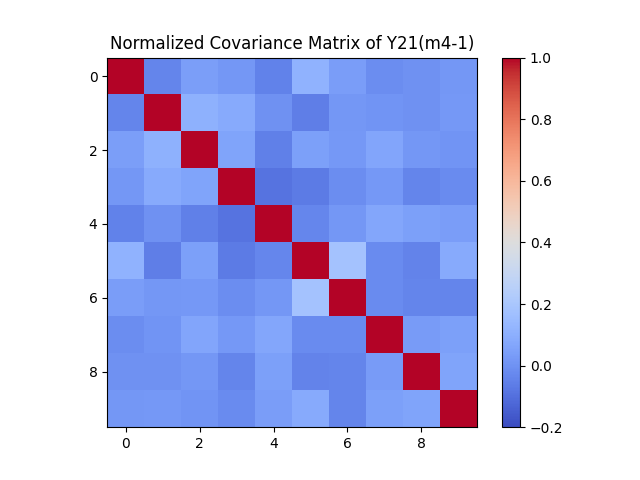}
 \caption{The normalized covariance matrix of the galaxy-galaxy lensing signal for `Y21(m4-1)'. The color map is the value of each element. }
\label{fig:cov}
\end{figure}

As described in Sec.~\ref{sec:method}, we measure the galaxy-galaxy lensing signal of samples and fit the signal with the model shown in Eq.~\ref{eq:model1}  and Eq.~\ref{eq:model2}. 
In the fitting, we use the likelihood as,
\begin{equation}
    {\rm ln}L=-0.5~({\rm Data - model})~(\rm{Cov}^{-1}) ~(\rm{ Data-model})^{T},
\end{equation}
where the Cov is the covariance matrix of the lensing signal, which accounts for the correlation between data points. The jackknife resampling method is used to estimate the covariance matrix of the lensing signal estimated via Eq.~\ref{eq:shearestimator}, which is computed as:
\begin{equation}
\rm Cov=\frac{K-1}{K}\sum_{k}^{K}\left(\Delta\Sigma_{(k)}-\Delta\Sigma_{(\cdot)}\right)^{T}\cdot\left(\Delta\Sigma_{(k)}-\Delta\Sigma_{(\cdot)}\right).
\end{equation}
This jackknife covariance estimates the variance by computing the deviations of each jackknife subsample
$\Delta\Sigma_{(\rm k)}$ from the mean $\Delta\Sigma_{(\cdot)}=\frac{1}{\rm K}\sum_{\rm k}\Delta\Sigma_{(\rm k)}$ scaling by $\frac{\rm K-1}{\rm K}$ to account for the bias introduced by leave-one-out resampling. 
In Fig.~\ref{fig:cov}, we show the normalized covariance matrix of signal error for ‘Y21(m4-1)’, as an example. We find that the covariances between different data points are quite weak.

The model has five free parameters ($M_{\rm vir}$, $c$, $R_{\rm mis}$, $p_{\rm cen}$ and $m$).
For the halo mass, $\log_{10}(M_{\rm vir})$, a flat prior is assigned to subsamples of Y21 within the range of [12.5, 15], while the flat prior for all subsamples of RM and Z21 is set to be within the range of [13, 15].
For all bins, a Gaussian prior is assumed for the concentration ($c$) based on the mass-concentration relation from \citet{shan2017mass}, as listed in Table.~\ref{tab:c_prior}. 
In addition, flat priors are applied to the miscentering length ($R_{\rm mis}$) in the range of [0, 0.5], and the probability of halos without miscentering effect ($p_{\rm cen}$) in the range of [0, 1]. Finally, a Gaussian prior of $0\pm 5$\% is set to the multiplicative bias, $m$.
We fit data using the MCMC method (\textsc{emcee}\footnote{\url{https://emcee.readthedocs.io/en/stable}}, \citealt{foreman2013emcee}). We employ 50 chains with an original length of $10,000$ steps, discarding the first $1,000$ steps to avoid the impact of the initial conditions (often referred to as burn-in).

\begin{figure*}
  \centering
\centering
  \begin{subfigure}{0.49\textwidth}
    \includegraphics[width=0.95\linewidth]{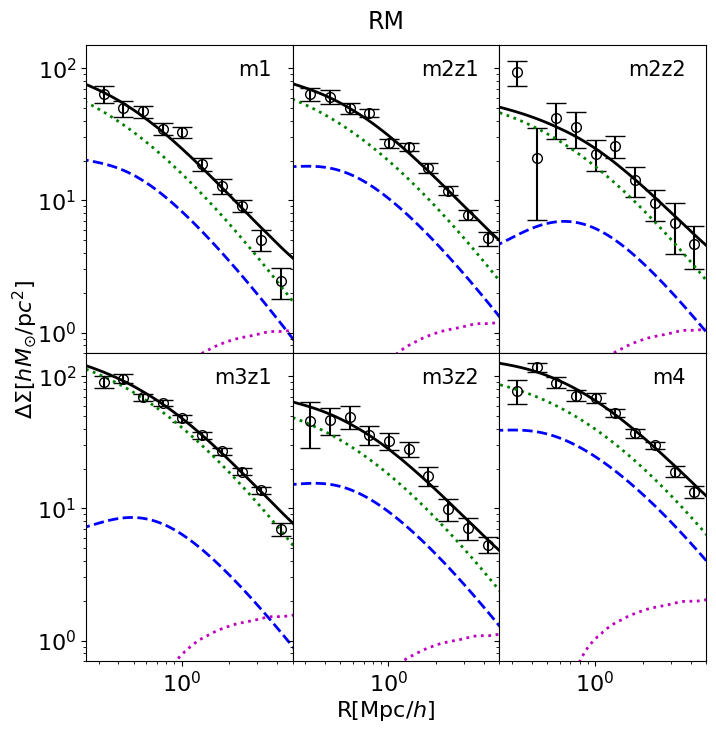}
    \label{fig:RMfit}
  \end{subfigure}
  \begin{subfigure}{0.49\textwidth}
    \includegraphics[width=1.05\linewidth]{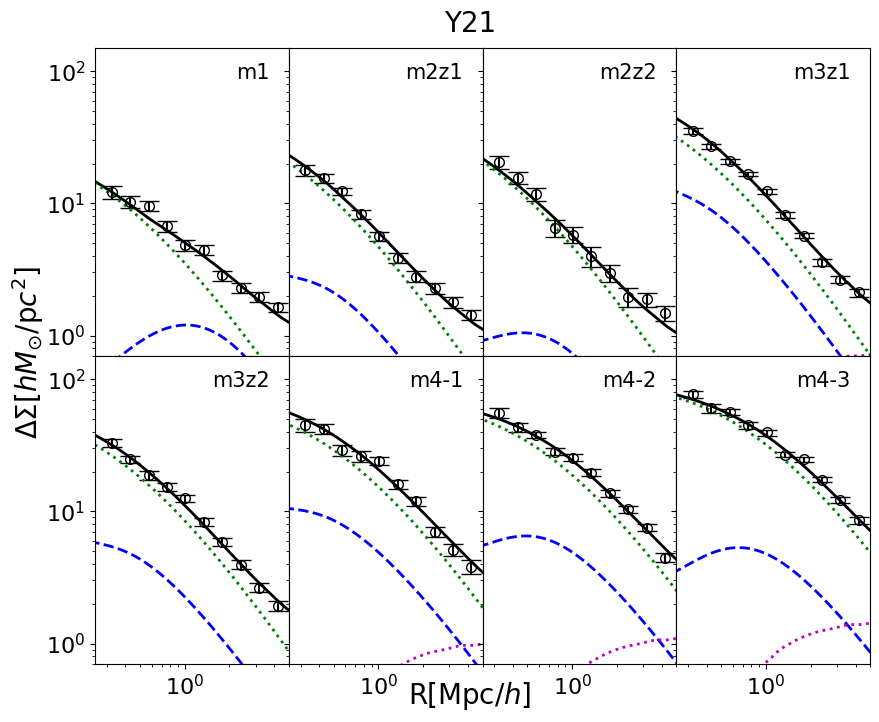}
    \label{fig:Yang21fit}
  \end{subfigure}
  \begin{subfigure}{0.49\textwidth}
    \includegraphics[width=1\linewidth]{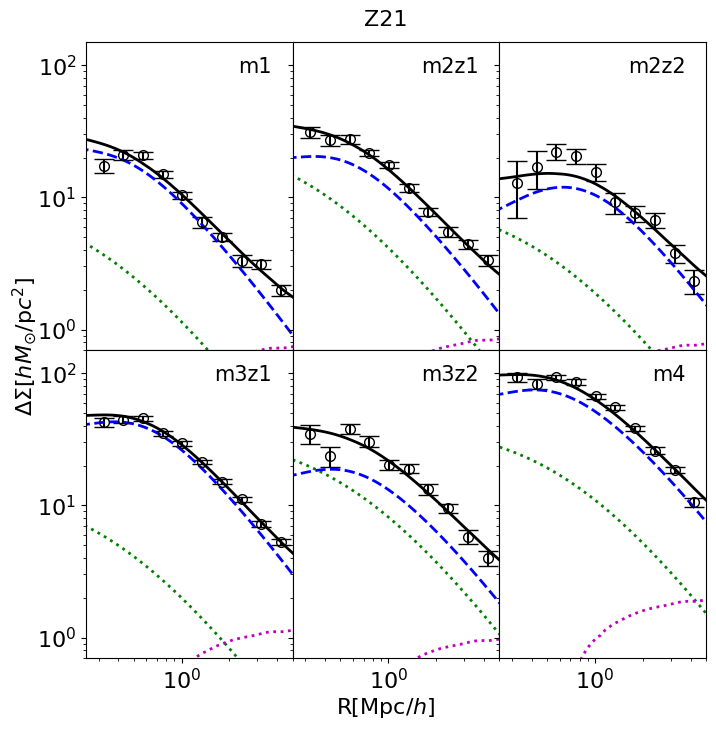}
    \label{fig:Zou21fit}
  \end{subfigure}
  \hfill
  \caption{Data and model-fitting for bins of RM, Y21, and Z21 catalogs. The hollow circles show the measured excess surface mass density $\Delta\Sigma\left(R\right)$, together with the errorbar for its 1$\sigma$ error. The best-fitted model is shown as the black curve, the contribution of the main dark matter halo as the green dotted curve, the miscentering term as the blue dashed curve, and the 2-halo term as the purple dotted curve.}
\label{fig:fit} 
\end{figure*}

The galaxy-galaxy lensing signals measured for the RM, Y21, and Z21 samples are displayed in Fig.~\ref{fig:fit}. As shown in Fig.~\ref{fig:distribution} and the first part of Table~\ref{tab:bin}, RM and Z21 are divided into six bins each, and Y21 is divided into eight bins. 
The optical fitting model is overlaid in Fig.~\ref{fig:fit}, and corresponding parameters are listed in the first part of Table~\ref{tab:result}. The model illustrates the contributions of the primary dark matter halos, the miscentering term, and the 2-halo term derived from the best-fit model. For RM and Y21 samples, the primary halo term dominates the total signal in most bins. For different sample bins, the influence of the miscentering term varies. The lensing-indicated center identification accuracy has a median value of 0.657 and 0.786 for RM and Y21, respectively, with the corresponding median value of center offset as 0.093 Mpc/$h$ and 0.108 Mpc/$h$. For Z21, the miscentering term dominates over the primary halo term, with a median center identification accuracy of 0.177, and a median center offset of 0.113 Mpc/$h$. For the sample of interest, the two-halo term rises beyond a radius of $\sim 1$ Mpc$/h$, but it does not contribute significantly to the total signal.

\begin{table*}
\centering
\caption{The optimal parameter fitting results.}
\begin{threeparttable}
\centering
\begin{tabular}{c|cccccc}
\hline\hline
Bin & log$_{10}$(M$_{\rm 200c,cat.}$) & log$_{10}$(M$_{\rm 200c,lens.}$) & $c$ & $R_{\rm mis}$[Mpc$/h$] & $p_{\rm cen}$& $m$\\
\hline
RM(m1)     & $13.988$ & $13.980_{-0.040}^{+0.041}$ & $4.808_{-0.622}^{+0.769}$ & $0.079_{-0.048}^{+0.105}$ & $0.660_{-0.403}^{+0.263}$ & $0.001_{-0.049}^{+0.050}$ \\
RM(m2$z$1) & $14.118$ & $14.142_{-0.037}^{+0.038}$ & $3.975_{-0.473}^{+0.650}$ & $0.093_{-0.054}^{+0.127}$ & $0.653_{-0.384}^{+0.252}$ & $0.002_{-0.050}^{+0.050}$ \\
RM(m2$z$2) & $14.145$ & $14.053_{-0.068}^{+0.072}$ & $2.927_{-0.546}^{+0.582}$ & $0.170_{-0.126}^{+0.198}$ & $0.711_{-0.337}^{+0.202}$ & $0.001_{-0.049}^{+0.050}$ \\
RM(m3$z$1) & $14.341$ & $14.415_{-0.034}^{+0.035}$ & $4.444_{-0.343}^{+0.521}$ & $0.129_{-0.108}^{+0.246}$ & $0.855_{-0.351}^{+0.106}$ & $0.003_{-0.050}^{+0.050}$ \\
RM(m3$z$2) & $14.345$ & $14.098_{-0.046}^{+0.045}$ & $3.321_{-0.420}^{+0.476}$ & $0.093_{-0.057}^{+0.126}$ & $0.651_{-0.404}^{+0.265}$ & $0.004_{-0.049}^{+0.049}$ \\
RM(m4)     & $14.679$ & $14.662_{-0.036}^{+0.036}$ & $3.485_{-0.302}^{+0.382}$ & $0.076_{-0.041}^{+0.108}$ & $0.614_{-0.398}^{+0.292}$ & $0.007_{-0.049}^{+0.050}$ \\
\hline
Y21(m1)       & $12.918$ & $13.184_{-0.092}^{+0.071}$ & $3.404_{-1.075}^{+1.068}$ & $0.385_{-0.127}^{+0.081}$ & $0.560_{-0.101}^{+0.140}$ & $0.010_{-0.049}^{+0.050}$ \\
Y21(m2$z$1)   & $13.230$ & $13.134_{-0.040}^{+0.042}$ & $3.825_{-0.557}^{+0.685}$ & $0.090_{-0.065}^{+0.290}$ & $0.817_{-0.434}^{+0.133}$ & $0.014_{-0.050}^{+0.050}$ \\
Y21(m2$z$2)   & $13.262$ & $13.120_{-0.043}^{+0.049}$ & $3.945_{-0.494}^{+0.523}$ & $0.163_{-0.140}^{+0.256}$ & $0.849_{-0.351}^{+0.107}$ & $0.010_{-0.049}^{+0.049}$ \\
Y21(m3$z$1)   & $13.592$ & $13.522_{-0.032}^{+0.033}$ & $4.899_{-0.488}^{+0.688}$ & $0.070_{-0.039}^{+0.074}$ & $0.682_{-0.385}^{+0.227}$ & $-0.005_{-0.050}^{+0.050}$ \\
Y21(m3$z$2)   & $13.604$ & $13.508_{-0.036}^{+0.037}$ & $3.855_{-0.367}^{+0.437}$ & $0.081_{-0.058}^{+0.147}$ & $0.792_{-0.418}^{+0.156}$ & $-0.006_{-0.050}^{+0.049}$ \\
Y21(m4-1)     & $14.089$ & $13.892_{-0.037}^{+0.039}$ & $3.365_{-0.315}^{+0.393}$ & $0.076_{-0.051}^{+0.157}$ & $0.758_{-0.431}^{+0.187}$ & $0.003_{-0.050}^{+0.049}$ \\
Y21(m4-2)     & $14.223$ & $14.016_{-0.036}^{+0.037}$ & $3.014_{-0.291}^{+0.415}$ & $0.126_{-0.093}^{+0.166}$ & $0.779_{-0.341}^{+0.152}$ & $0.000_{-0.050}^{+0.050}$ \\
Y21(m4-3)     & $14.456$ & $14.316_{-0.035}^{+0.036}$ & $2.922_{-0.221}^{+0.366}$ & $0.162_{-0.139}^{+0.224}$ & $0.850_{-0.307}^{+0.106}$ & $0.004_{-0.049}^{+0.049}$ \\
\hline
Z21(m1) & $13.890$ & $13.468_{-0.037}^{+0.037}$ & $4.045_{-0.459}^{+0.487}$ & $0.085_{-0.012}^{+0.012}$ & $0.113_{-0.083}^{+0.157}$ & $0.002_{-0.049}^{+0.050}$ \\
Z21(m2$z$1) & $14.060$ & $13.753_{-0.037}^{+0.037}$ & $4.124_{-0.705}^{+0.808}$ & $0.110_{-0.023}^{+0.029}$ & $0.254_{-0.165}^{+0.198}$ & $0.002_{-0.050}^{+0.049}$ \\
Z21(m2$z$2) & $14.066$ & $13.699_{-0.048}^{+0.046}$ & $2.998_{-0.537}^{+0.564}$ & $0.128_{-0.028}^{+0.036}$ & $0.184_{-0.133}^{+0.239}$ & $-0.002_{-0.049}^{+0.049}$ \\
Z21(m3$z$1) & $14.244$ & $14.099_{-0.030}^{+0.031}$ & $5.330_{-0.562}^{+0.642}$ & $0.118_{-0.012}^{+0.013}$ & $0.067_{-0.048}^{+0.077}$ & $0.003_{-0.050}^{+0.050}$ \\
Z21(m3$z$2) & $14.244$ & $13.958_{-0.042}^{+0.041}$ & $3.054_{-0.485}^{+0.522}$ & $0.115_{-0.034}^{+0.053}$ & $0.368_{-0.237}^{+0.275}$ & $0.000_{-0.049}^{+0.050}$ \\
Z21(m4) & $14.567$ & $14.616_{-0.031}^{+0.032}$ & $4.431_{-0.376}^{+0.403}$ & $0.104_{-0.015}^{+0.018}$ & $0.169_{-0.117}^{+0.157}$ & $0.019_{-0.049}^{+0.049}$ \\
\hline
\hline
RM(cen.)       & $14.195$ & $14.239_{-0.033}^{+0.034}$ & $4.962_{-0.560}^{+0.831}$ & $0.124_{-0.077}^{+0.092}$ & $0.698_{-0.295}^{+0.198}$ & $0.001_{-0.050}^{+0.049}$ \\
Y21(cen.)      & $14.146$ & $14.237_{-0.032}^{+0.034}$ & $4.743_{-0.421}^{+0.686}$ & $0.120_{-0.090}^{+0.133}$ & $0.781_{-0.316}^{+0.158}$ & $0.001_{-0.049}^{+0.049}$ \\
RM(mem. sub1)  & $14.196$ & $14.235_{-0.033}^{+0.034}$ & $4.924_{-0.395}^{+0.546}$ & $0.085_{-0.063}^{+0.170}$ & $0.802_{-0.403}^{+0.150}$ & $0.003_{-0.050}^{+0.050}$ \\
Y21(mem. sub1) & $14.183$ & $14.238_{-0.031}^{+0.033}$ & $5.022_{-0.391}^{+0.518}$ & $0.074_{-0.056}^{+0.182}$ & $0.808_{-0.428}^{+0.149}$ & $0.003_{-0.048}^{+0.048}$ \\
RM(mem. sub2)  & $14.363$ & $14.381_{-0.044}^{+0.046}$ & $5.172_{-0.643}^{+0.728}$ & $0.081_{-0.061}^{+0.184}$ & $0.798_{-0.420}^{+0.159}$ & $0.006_{-0.049}^{+0.049}$ \\
Y21(mem. sub2) & $14.410$ & $14.378_{-0.049}^{+0.049}$ & $4.979_{-0.759}^{+0.855}$ & $0.105_{-0.038}^{+0.048}$ & $0.369_{-0.257}^{+0.362}$ & $0.003_{-0.049}^{+0.050}$ \\
RM(un.)        & $14.200$ & $14.183_{-0.041}^{+0.043}$ & $3.605_{-0.420}^{+0.668}$ & $0.168_{-0.132}^{+0.182}$ & $0.786_{-0.263}^{+0.147}$ & $0.001_{-0.050}^{+0.050}$ \\
Y21(un.)       & $14.149$ & $14.001_{-0.037}^{+0.038}$ & $3.009_{-0.330}^{+0.466}$ & $0.173_{-0.138}^{+0.156}$ & $0.796_{-0.256}^{+0.137}$ & $-0.001_{-0.050}^{+0.050}$ \\
\hline
\hline
\end{tabular}
\begin{tablenotes}[flushleft] 
\item \textbf{Note}: The first two columns list the bin name and the average mass provided by lens catalog. The following five columns show the fitting results of the halo mass, concentration, miscentering length, fraction of halos without miscentering, and the multiplicative bias. Both M$_{\rm 200c,cat.}$ and M$_{\rm 200c,lens.}$ are in the unit of $M_{\rm \odot}h^{-1}$.
\end{tablenotes}
     \end{threeparttable}
\label{tab:result} 
\end{table*}

\subsection{Comparison of halo masses from catalogs and lensing signals}
\label{sec:Mcat_Mlens}

We compare the halo masses estimated from the group catalogs with those derived from the fitting of galaxy-galaxy lensing signals. As described in Sec.~\ref{sec:lens}, the halo masses in the three group catalogs are estimated using different methods. 
In the RM catalog, the group mass is estimated through mass-richness relation \citep{mcclintock2019dark,to2021dark,costanzi2021cosmological} as 
 \begin{align}
M(\lambda, z) \equiv \langle M \mid \lambda, z \rangle = M_0 \left( \frac{\lambda}{\lambda_0} \right)^{F_\lambda} \left( \frac{1 + z}{1 + z_0} \right)^{G_z}.
\end{align}
 Using pivot values \( \lambda_0 = 40 \) and \( z_0 = 0.35 \), the best-fit parameters reported are \( \log_{10} M_0 = 14.489 \pm 0.011 \pm 0.019 \), \( F_\lambda = 1.356 \pm 0.051 \pm 0.008 \), and \( G_z = -0.30 \pm 0.30 \pm 0.06 \). These values are obtained from Table.~4 of \cite{mcclintock2019dark}, whose uncertainties refer to the 68\% confidence intervals including the statistical (first-term) error and systematic (second-term) error.

Y21 catalog adopts an abundance matching method to estimate the group mass. The group masses are assigned via abundance matching under Planck18 cosmology \citep{collaboration2020planck}, making them cosmology-dependent \citep{wang2022halo}.
The mass of Z21 groups is estimated by the group luminosity and redshift \citep{zou2021galaxy}, following the relation:
\begin{align}
\log\bigl(M_{\rm 500}\bigr)=A_1\log\bigl(L_{\rm 1\text{Mpc}}\bigr)+A_2\log\bigl(1+z\bigr)+A_3 \,,
\end{align}
where $z$ denotes the redshift, and $L_{\rm 1Mpc}$ (i.e., the group luminosity) represents the total \textit{r}-band luminosity of galaxies within a 1 Mpc radius from the central galaxy. The model parameters, $A_1 = 0.81 \pm 0.02$, $A_2 = 0.50 \pm 0.14$, and $A_3 = 12.61 \pm 0.04$, are calibrated using X-ray and SZ observations with different depths.

\begin{figure*}
\centering
\includegraphics[width=0.8\textwidth]{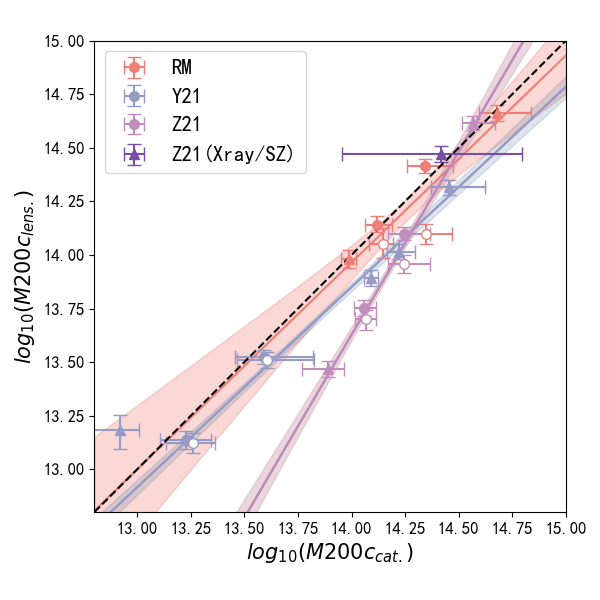}
\caption{The relation between the lensing mass and the estimated mass provided by catalogs for the RM bins (red), Y21 bins (blue), the Z21 bins (purple), and the `Z21(X-ray/SZ)' sample (dark purple). The Y-axis represents the lensing mass, while the X-axis represents the estimated mass from the halo catalog. Both masses are in the unit of $M_{\rm \odot}h^{-1}$. Solid points represent low-redshift bins (m2$z$1, m3$z$1), hollow points represent high-redshift bins (m2$z$2, m3$z$2), and triangular symbols represent the bins without redshift restriction (m1, m4). The vertical error bars indicate the 1$\sigma$ uncertainty of the lensing fitted halo mass. The horizontal error bars represent the standard deviation of mass from the halo catalog. The black dashed line represents the 1:1 relation. Solid lines with light shade regions in different colors are the best fitting relations with $1\sigma$ uncertainty for the corresponding catalogs. In the fitting, `Y21(m1)' and `Z21(X-ray/SZ)' data are not taken into account.}
\label{fig:lensest} 
\end{figure*}

The mass of groups in three catalogs are all converted to $M_{200c}$ using the NFW model with the \textsc{Colossus} package\footnote{\url{https://bdiemer.bitbucket.io/colossus/}} \citep{Diemer2018}.
The comparison of halo mass from the halo catalog ($M_{\rm 200c,cat.}$) and from the lensing signal ($M_{\rm 200c,lens.}$) is shown in Fig.~\ref{fig:lensest}.
Overall, the catalog mass correlates linearly with the lensing mass in all cases. We perform a linear fit for the relation between the mass proxy of three group catalogs and the lensing mass, respectively. For the linear fit, we use 6 bins for RM and Z21, and 7 bins for Y21. The `Y21(m1)' data is excluded for the obvious deviation from the linear relation, which likely comes from the limitation of lensing measurements for less-massive halos. We adopt the form of the linear relation as
\begin{align}
\log_{10}({\rm M_{ 200c,lens.}}) = A\log_{10}({\rm M_{200c,cat.}})+B.
    \label{eq:m_m}
\end{align}
The best fitting parameters of $A$ and $B$ are shown in Table~\ref{tab:Table5}.

\begin{table}
\caption{Best-fitting parameters of Eq.~\ref{eq:m_m}.}
\centering
\begin{tabular}{c|cc}
\hline\hline
Catalog & $A$ & $B$\\
\hline
RM  & 0.97 &0.45\\
Y21 & 0.93 &0.77\\
Z21 & 1.71 &-10.29 \\
\hline
\hline
\end{tabular}
\label{tab:Table5} 
\end{table}

The median of ${\rm log_{10}(M_{200c,cat.})-log_{10}(M_{200c,lens.})}$ of different mass bins, is 0.013 for RM, 0.118 for Y21, and 0.297 for Z21. Among these three catalogs, the mass estimates provided by RM show the best agreement with galaxy-galaxy lensing measurements. Notably, RM's mass estimates in low-redshift bins align more closely with lensing results compared to high-redshift bins, indicating the need for redshift-dependent mass calibration.

Y21 is the only catalog among the three that includes estimates for low-mass groups, and provides reasonably good agreement with lensing results, except for the lowest mass bin. Overall, Y21 systematically provides relatively higher mass estimates than those inferred from lensing, and is consistent with \citet{sun2022cross}, where group masses assigned from luminosity are found to be systematically higher than the true halo masses in mocks, with a typical scatter of $\sim$ 0.3 dex and a bias increasing at lower masses. Additionally, when comparing two redshift bins with similar group masses, we find that the lensing mass tends to be slightly lower at higher redshift, a trend also noted in \citet{wang2022halo}.

The Z21 group catalog consistently overestimates halo masses compared to lensing measurements, with the degree of overestimation varying across redshift bins. The absence of a boost factor correction may result in a slight bias toward lower lensing mass. Interestingly, the subset of Z21 calibrated using X-ray/SZ data aligns well with lensing results, suggesting a potential selection bias in the calibration subset relative to the full dataset. The `Z21(X-ray/SZ)' sample predominantly consists of massive groups, effectively calibrating the high-mass bin but offering limited calibration for lower-mass halos.

\begin{figure*}
\centering
  \begin{subfigure}{0.48\textwidth}
    \includegraphics[width=1\linewidth]{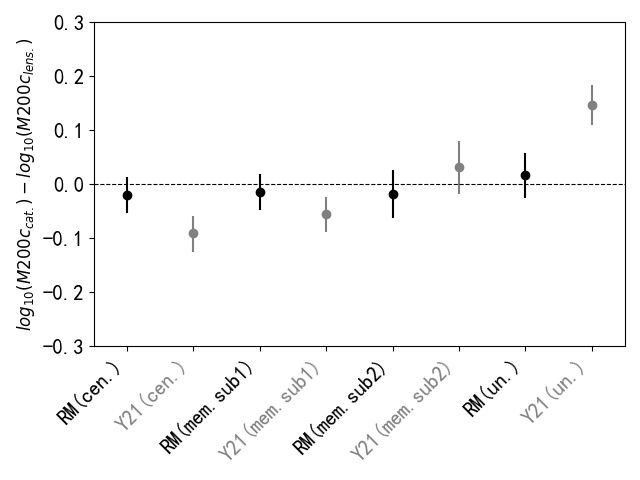}
  \end{subfigure}
  \hfill
  \begin{subfigure}{0.48\textwidth}
    \includegraphics[width=1\linewidth]{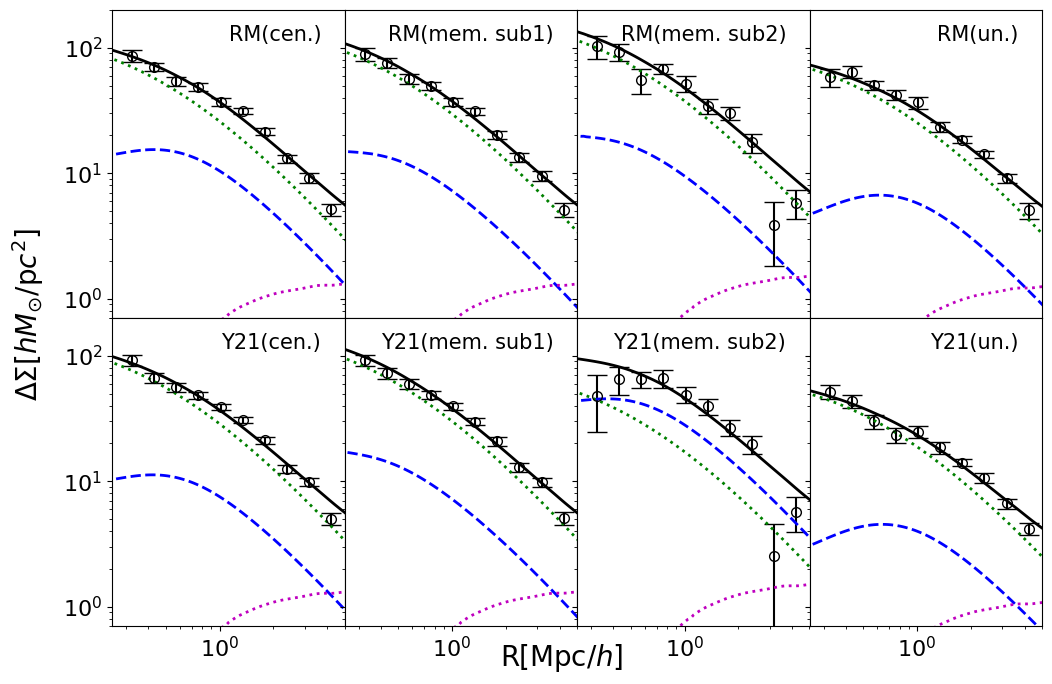}
  \end{subfigure}
\caption{Left panel is the difference (with error bar) between the estimated mass from the halo catalog and the lensing mass for cross-matched and unmatched samples. The right panels show the fitting details for the eight samples shown in the left panel. The symbols are the same with Fig.~\ref{fig:fit}.}
\label{fig:yrmc_new} 
\end{figure*}

We further compare the lensing measurements with the mass estimates from the RM and Y21 catalogs for the cross-matched samples, as shown in the left panel of Fig.~\ref{fig:yrmc_new}. In the right panel of Fig.~\ref{fig:yrmc_new}, we show the measurement and fitting of the lensing signal of all these cross-matched samples, overlaid with the optical fitting models and components.
Besides the cross-matched samples described in Sec.~\ref{sec:crosscompare_RM_Y21}, we also check the unique groups identified only in the RM or Y21, which are out of the cross-matching. To exclude the limitation of the survey coverage, we only consider groups within the main overlapping region, with RA within [140, 240]~deg and DEC within [0, 60]~deg. We select only massive Y21 groups with the halo mass $>10^{14} M_{\odot}/h$, and make no selection on RM groups. These clusters not included in the aforementioned cross-matched samples are named as `RM(un.)' and `Y21(un.)', including 7,920 and 33,383 groups, respectively.

In the left panel of Fig.~\ref{fig:yrmc_new}, we find that RM makes a more accurate estimation of halo mass than Y21 across all types of cross-matching samples, with the halo mass differences of $<0.05$ dex.
However, the Y21 groups fall into two main categories.
For Y21 groups sharing the central galaxy with RM groups, i.e., both `Y21(cen.)' sample and `Y21(mem. sub1)' sample, Y21 tends to make an under-estimation of halo mass. Otherwise, Y21 makes a systematic mass over-estimation, for both `Y21(mem. sub2)' sample and`Y21(un.)' sample. In the right panel of Fig.~\ref{fig:yrmc_new}, we find that the miscentering term of `Y21(mem. sub2)' is more significant compared to the center match scenario (the `Y21(mem. sub1)' sample). Thus, the stronger projection effect \citep{Zu2017} and miscentering effect likely introduce Y21 groups a systematic bias in the abundance-matching method.

\section{Conclusion}
\label{sec:conclusion}

In this work, we compare the halo masses of galaxy groups identified in the RM, Y21, and Z21 catalogs, which are based on different methods and datasets. Using the galaxy-galaxy lensing signal from DECaLS DR8, we measure the projected excess surface density profiles of halo subsamples in specific redshift and mass bins. These profiles are modeled as a combination of contributions from the central halo, the miscentering effect, and neighboring halos.

A linear correlation is observed between the catalog-estimated halo masses ($M_{\rm 200c,cat.}$) and those derived from lensing signals ($M_{\rm 200c,lens.}$). RM shows the best agreement, especially for lower-redshift groups, with minor deviations in higher-redshift bins. 
Y21 is the only catalog covering the low-mass regime, yielding mass estimates broadly consistent with lensing, except for the lowest mass bin. Z21 systematically overestimates halo mass compared to the lensing result, although their mass estimation of the calibration sample matches the corresponding lensing result. This discrepancy may partly stem from the lower representation of less massive halos in the calibration sample, as well as a possible mass under-estimation using the lensing method due to the absence of the boost factor correction.

Comparing the halo mass of the cross-matched samples from RM and Y21, we find that RM makes a consistent mass estimation with the lensing result, regardless of whether the halos are also identified in the Y21 catalog or not.
However, the mass estimation in the cross-matched samples from Y21 varies with different cross-matching results with RM. The Y21 tends to underestimate the halo mass for groups with the same center identified by RM, and over-estimates the halo mass for groups with different center identification or uniquely identified groups in the Y21 catalog. The latter likely comes from their stronger projection effects and miscentering effect, which likely bias the abundance-matching method. This result emphasizes the importance of careful study of subsamples with different properties in group catalogs for a better understanding of halos, such as halo mass estimation.

Our results highlight that the choice of galaxy group catalog has a critical impact on halo mass estimates. Systematic differences in group finding, center identification, and mass calibration can propagate into the modeling of the excess surface density profiles and further bias the constraints on halo occupation and matter clustering (e.g., \citealt{mandelbaum2006density,johnston2007cross}). For cluster astrophysics, these differences affect derived scaling relations, such as the mass–richness or mass–luminosity relation, which are widely used to infer cluster physical properties and their baryonic content (e.g., \citealt{rozo2009improvement,rykoff2014redmapper,Xu2021,xu2024measurement}).

In the context of cluster cosmology, biased halo masses directly translate into biased estimates of cluster abundance and mass function, which are key probes for constraining cosmological parameters (e.g., \citealt{vikhlinin2009chandra,planck2016results,wu2022optical,zhang2022effect,xu2023desi}). Our findings stress that robust cosmological constraints from cluster counts or stacked lensing require careful cross-checks of group finding and centering determination, and highlight the importance of combining multiple independent catalogs and lensing datasets to calibrate and mitigate systematics.

Future large surveys such as the Legacy Survey of Space and Time (LSST) of Rubin Observatory, the Euclid survey, and the Chinese Survey Space Telescope (CSST) will greatly improve the statistical power of lensing and cluster cosmology analyses. To fully exploit these datasets, it is essential to understand and quantify catalog-dependent systematics like those identified in this study. This will enable more precise and unbiased measurements of the dark matter distribution, galaxy evolution in group environments, and fundamental cosmological parameters.

\appendix
\section{Prior of Concentration}
In Table.~\ref{tab:c_prior}, the prior of concentration for each bin is listed.

\begin{table}[ht]
\centering
\caption{The Gaussian priors of the concentration.}
\centering
\begin{threeparttable}
\begin{tabular}{c c | c c | c c | c c}
\hline\hline
Bin.      & $c$    & Bin.      & $c$  & Bin.      & $c$  & Bin.      & $c$\\     
\hline
RM(m1)     &  [2.69, 1.20] & Y21(m1)    &   [3.91, 1.01] &  Z21(m1)     & [2.94, 0.63] & RM(cen.)       & [2.44, 1.20] \\  
RM(m2$z$1) &  [2.55, 1.22] & Y21(m2$z$1)&   [3.33, 1.06] &  Z21(m2$z$1) & [2.54, 1.19] & Y21(cen.)      & [2.46, 1.20] \\  
RM(m2$z$2) &  [2.76, 0.66] & Y21(m2$z$2)&   [3.72, 0.59] &  Z21(m2$z$2) & [2.71, 0.63] & RM(mem. sub1)  & [2.42, 1.20] \\  
RM(m3$z$1) &  [2.38, 1.25] & Y21(m3$z$1)&   [2.99, 1.14] &  Z21(m3$z$1) & [2.40, 1.21] & Y21(mem. sub1) & [2.43, 1.20] \\  
RM(m3$z$2) &  [2.54, 0.66] & Y21(m3$z$2)&   [3.30, 0.62] &  Z21(m3$z$2) & [2.55, 0.64] & RM(mem. sub2)  & [2.44, 1.22] \\  
RM(m4)     &  [2.28, 0.68] & Y21(m4-1)  &   [2.67, 0.63] &  Z21(m4)     & [2.28, 0.65] & Y21(mem. sub2) & [2.51, 1.20] \\ 
~          &           ~   & Y21(m4-2)  &   [2.54, 0.63] &  ~           &  ~           &  RM(un.)       & [2.40, 1.20] \\  
~          &           ~   & Y21(m4-3)  &   [2.37, 0.65] &  ~           &  ~           &  Y21(un.)      & [2.54, 0.62] \\ 
\hline
\hline
\end{tabular}
\begin{tablenotes} 
\item \textbf{Note}:
The mean and the full width at half maximum (FWHM) of the Gaussian prior of concentration are listed for each bin.
\end{tablenotes}
\end{threeparttable}
\label{tab:c_prior}
\end{table}

\section{Boost factor}
\label{app:boost}

\begin{figure}
    \centering
       \includegraphics[width=0.6\textwidth]{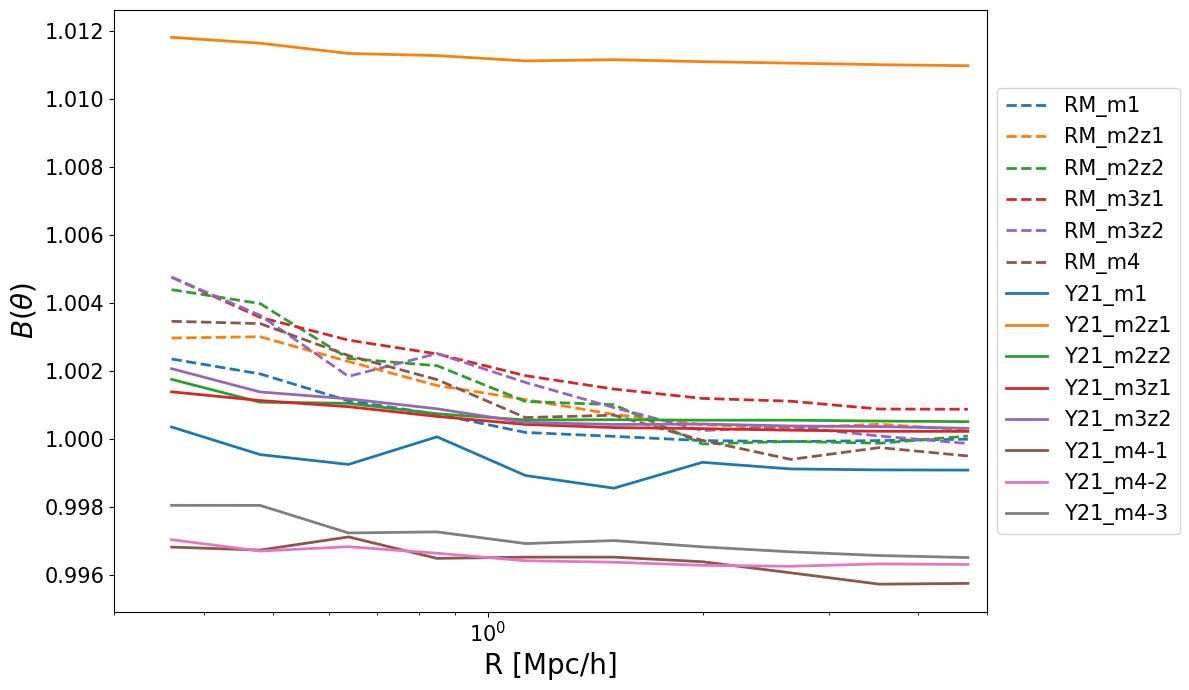}
\caption{The boost factor of bins of RM and Y21 catalogs.}
\label{fig:boost}
\end{figure}

In our analysis, the boost factor as a function of radius is calculated by Eq.~\ref{eq:boost} for each bins of RM and Y21, and shown in Fig.~\ref{fig:boost}. The results indicate a small effect of contamination from physically associated pairs, especially at small scales. The boost factors for some bins are smaller than 1, which might come from the limitation of the simulation data. This bias could be absorbed into the multiplicative bias term. Therefore, it does not bias our mass estimation. 

Besides, the Z21 catalog has no random catalog, but its boost factors are expected to be smaller than Y21. The main reasons include their large value of mass and richness, which result in photometric redshifts with small boost factors. In addition, the photometric redshifts in Y21 and Z21 are estimated with different methods, which makes it difficult to make an accurate prediction. And the detailed discussion about the comparison of these photometric redshift methods is out of the scope of this work. 

\section{Multiplicative Bias}
\label{app:multiplicative}

\begin{figure}
  \centering
    \includegraphics[width=0.5\linewidth]{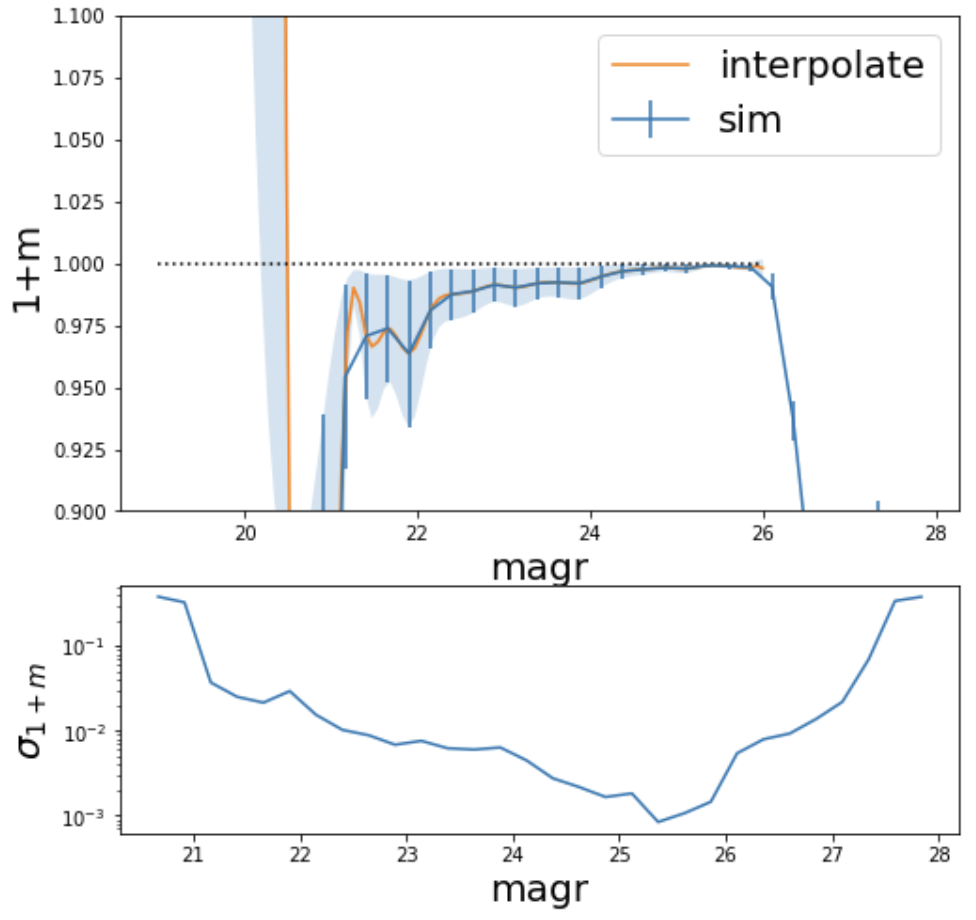}
 \caption{The top panel is the multiplicative bias and its error of shear catalog from DECaLS DR9 at different $r$-band magnitudes. The `sim' data presents the measured values of $1 + m$ across $r$-band magnitude bins. The `interpolate' curve corresponds to the prediction of the $1 + m$ values from the interpolation of a neural network model. The bottom panel is the residuals of the measured data.}
\label{fig:shear_bias}
\end{figure}

The multiplicative bias, $\mathsf{A}_m = 1 + m$, is analyzed with respect to $r$-band magnitude using DECaLS DR9 data to evaluate its impact on the shear measurement. Fig.~\ref{fig:shear_bias} presents the measured values of $1 + m$ across magnitude bins after the calibration in the $z$-band, along with the residuals. The prediction of 1+$m$ across the $r$-band is also shown in Fig.~\ref{fig:shear_bias}, which is derived from a neural network model trained to interpolate, providing a smooth estimate of the trend.
In addition, the shear catalog of DECaLS DR8 calibrated with $z$-band magnitude and galaxy size is used in our analysis. Thus, the residual shown in Fig.~\ref{fig:shear_bias}, which comes from the measurement with the DECaLS DR9 shear catalog after $z$-band calibration, is a rough estimation.

\software{
\textsc{cluster~toolkit} \citep{Smith2003,Eisenstein1998,Takahashi2012}, 
\textsc{colossus} software \citep{Diemer2018}, 
\textsc{camb} \citep{Challinor2011,Lewis1999}, 
\textsc{swot} \citep{Coupon2012}, 
\textsc{emcee} \citep{emcee}.}

\begin{acknowledgements}
We acknowledge the support by National Key R$\&$D Program of China No. 2022YFF0503403, the support of National Nature Science Foundation of China (Nos 11988101,12022306, 12203063, 12203084), the support from the Ministry of Science and Technology of China (Nos. 2020SKA0110100), the science research grants from the China Manned Space Project (Nos. CMS-CSST-2025-A03, CMS-CSST-2021-B01, CMS-CSST-2021-A01), CAS Project for Young Scientists in Basic Research (No. YSBR-062), and the support from K.C.Wong Education Foundation. HYS acknowledges the support from NSFC of China under grant 11973070, Key Research Program of Frontier Sciences, CAS, Grant No. ZDBS-LY-7013 and Program of Shanghai Academic/Technology Research Leader. 

The DESI Legacy Imaging Surveys consist of three individual and complementary projects: the Dark Energy Camera Legacy Survey (DECaLS), the Beijing-Arizona Sky Survey (BASS), and the Mayall z-band Legacy Survey (MzLS). DECaLS, BASS and MzLS together include data obtained, respectively, at the Blanco telescope, Cerro Tololo Inter-American Observatory, NSF’s NOIRLab; the Bok telescope, Steward Observatory, University of Arizona; and the Mayall telescope, Kitt Peak National Observatory, NOIRLab. NOIRLab is operated by the Association of Universities for Research in Astronomy (AURA) under a cooperative agreement with the National Science Foundation. Pipeline processing and analyses of the data were supported by NOIRLab and the Lawrence Berkeley National Laboratory (LBNL). Legacy Surveys also uses data products from the Near-Earth Object Wide-field Infrared Survey Explorer (NEOWISE), a project of the Jet Propulsion Laboratory/California Institute of Technology, funded by the National Aeronautics and Space Administration. Legacy Surveys was supported by: the Director, Office of Science, Office of High Energy Physics of the U.S. Department of Energy; the National Energy Research Scientific Computing Center, a DOE Office of Science User Facility; the U.S. National Science Foundation, Division of Astronomical Sciences; the National Astronomical Observatories of China, the Chinese Academy of Sciences and the Chinese National Natural Science Foundation. LBNL is managed by the Regents of the University of California under contract to the U.S. Department of Energy. The complete acknowledgments can be found at https://www.legacysurvey.org/acknowledgment/.
\end{acknowledgements}

\bibliography{main} 
\bibliographystyle{aasjournal}

\end{document}